\documentclass{article}

\usepackage[utf8]{inputenc}
\usepackage[T1]{fontenc}    

\usepackage{booktabs}       
\usepackage{amsfonts}       
\usepackage{nicefrac}       
\usepackage{microtype}      
\usepackage{lipsum}

\usepackage{tikz,ifthen,xstring,calc,pgfkeys,pgfopts}
\usepackage{macros/tikz-uml}
\usepackage{courier}

\usepackage{multirow}
\usepackage{array}
\usepackage{rotating}
\usepackage{makecell}

\usepackage{hyperref}
\usepackage{url}
\usepackage{tcolorbox}

\usepackage{algorithm,algpseudocode}
\usepackage{listings}

\usepackage{mathtools}
\usepackage{amsmath,amssymb,amsthm}
\usepackage{thmtools}

\usepackage{graphicx}
\usepackage{adjustbox}

\usepackage{todonotes}

\usepackage[charter]{mathdesign}
\usepackage[mathcal]{eucal}
\usepackage{bbm}

\usepackage{color, colortbl}
\definecolor{Gray}{gray}{0.9}
\definecolor{LightCyan}{rgb}{0.88,1,1}

\usepackage[shortlabels]{enumitem}
\usepackage{scrextend}
\addtokomafont{labelinglabel}{\sffamily}

\usepackage[numbers]{natbib}

\newcommand{\id}{\operatorname{id}}

\newcommand{\card}{\#}

\newcommand{\calC}{\mathcal{C}}

\newcommand{\calN}{\mathcal{N}}

\newcommand{\calX}{\mathcal{X}}
\newcommand{\calY}{\mathcal{Y}}

\newcommand{\EE}{\ensuremath{\mathbb{E}}}

\newcommand{\NN}{\ensuremath{\mathbb{N}}}

\newcommand{\RR}{\ensuremath{\mathbb{R}}}

\newcommand{\ZZ}{\ensuremath{\mathbb{Z}}}

\newcommand{\OOne}{\ensuremath{\mathbbm{1}}}

\makeatletter
\providecommand*{\diff}%
        {\@ifnextchar^{\DIfF}{\DIfF^{}}}
\def\DIfF^#1{%
        \mathop{\mathrm{\mathstrut d}}%
                \nolimits^{#1}\gobblespace
}
\def\gobblespace{%
        \futurelet\diffarg\opspace}
\def\opspace{%
        \let\DiffSpace\!%
        \ifx\diffarg(%
                \let\DiffSpace\relax
        \else
                \ifx\diffarg\[%
                        \let\DiffSpace\relax
                \else
                        \ifx\diffarg\{%
                                \let\DiffSpace\relax
                        \fi\fi\fi\DiffSpace}
\makeatother

\newcommand{\simiid}{\overset{\mbox{\tiny i.i.d.}}{\sim}}

\usepackage{macros/arxiv}
\usepackage{newfloat}
\usepackage{chngcntr}
\usepackage{caption}
\usepackage{tabularx}
\usepackage{booktabs}       
\setlist[description]{leftmargin=0.5cm}
\usepackage{xcolor}
\usepackage{graphicx}

\definecolor{codegreen}{rgb}{0,0.6,0}
\definecolor{codegray}{rgb}{0.5,0.5,0.5}
\definecolor{codepurple}{rgb}{0.58,0,0.82}
\definecolor{backcolour}{rgb}{0.95,0.95,0.92}

\usepackage{listings}

\definecolor{halfgray}{gray}{0.55}
\definecolor{iPython_frame}{RGB}{207, 207, 207}
\definecolor{iPython_bg}{RGB}{247, 247, 247}
\definecolor{iPython_red}{RGB}{186, 33, 33}
\definecolor{iPython_green}{RGB}{0, 128, 0}
\definecolor{iPython_cyan}{RGB}{64, 128, 128}
\definecolor{iPython_purple}{RGB}{170, 34, 255}
\lstdefinestyle{iPython}{
	language=Python,
	frame=lines,
	backgroundcolor=\color{white},
	commentstyle=\color{iPython_cyan},
	keywordstyle=\color{iPython_green},
	numberstyle=\ttfamily\tiny\color{halfgray},
	escapechar=\¢,escapebegin=\color{iPython_green},
	stringstyle=\color{iPython_red},
	basicstyle=\ttfamily\footnotesize,
	breakatwhitespace=false,
	breaklines=true,
	captionpos=t,
	keepspaces=true,
	numbers=left,
	numbersep=5pt,
	showspaces=false,
	showstringspaces=false,
	showtabs=false,
	tabsize=2,
	morekeywords={access,and,break,class,continue,def,del,elif,else,except,exec,finally,for,from,global,if,import,in,is,lambda,not,or,pass,print,raise,return,try,while},%
	morekeywords=[2]{abs,all,any,basestring,bin,bool,bytearray,callable,chr,classmethod,cmp,compile,complex,delattr,dict,dir,divmod,enumerate,eval,execfile,file,filter,float,format,frozenset,getattr,globals,hasattr,hash,help,hex,id,input,int,isinstance,issubclass,iter,len,list,locals,long,map,max,memoryview,min,next,object,oct,open,ord,pow,property,range,raw_input,reduce,reload,repr,reversed,round,set,setattr,slice,sorted,staticmethod,str,sum,super,tuple,type,unichr,unicode,vars,xrange,zip,apply,buffer,coerce,intern},%
	sensitive=true,%
	morecomment=[l]\#,%
	morestring=[b]',%
	morestring=[b]",%
	morestring=[s]{'''}{'''},
	morestring=[s]{"""}{"""},
	morestring=[s]{r'}{'},
	morestring=[s]{r"}{"},%
	morestring=[s]{r'''}{'''},%
	morestring=[s]{r"""}{"""},%
	morestring=[s]{u'}{'},
	morestring=[s]{u"}{"},%
	morestring=[s]{u'''}{'''},%
	morestring=[s]{u"""}{"""},%
	literate=
	*{+}{{{\color{iPython_purple}+}}}1
	{-}{{{\color{iPython_purple}-}}}1
	{*}{{{\color{iPython_purple}$^\ast$}}}1
	{/}{{{\color{iPython_purple}/}}}1
	{^}{{{\color{iPython_purple}\^{}}}}1
	{?}{{{\color{iPython_purple}?}}}1
	{!}{{{\color{iPython_purple}!}}}1
	{\%}{{{\color{iPython_purple}\%}}}1
	{<}{{{\color{iPython_purple}<}}}1
	{>}{{{\color{iPython_purple}>}}}1
	{|}{{{\color{iPython_purple}|}}}1
	{\&}{{{\color{iPython_purple}\&}}}1
	{~}{{{\color{iPython_purple}~}}}1
	{=}{{{\color{iPython_purple}=}}}1
	{==}{{{\color{iPython_purple}==}}}2
	{<=}{{{\color{iPython_purple}<=}}}2
	{>=}{{{\color{iPython_purple}>=}}}2
	{+=}{{{+=}}}2
	{-=}{{{-=}}}2
	{*=}{{{$^\ast$=}}}2
	{/=}{{{/=}}}2,
}
\newcommand{\pattern}[1]{
	\textsc{#1}
}

\newcommand{\software}[1]{
	\texttt{#1}
}
\newcommand{\litem}[1]{
	\item{\bfseries #1}
}

\DeclareFloatingEnvironment[
fileext=los,
listname={List of Scientific Typing Schemas},
name=Scitype,
placement=t,
]{scitype}
\counterwithout{scitype}{section}

\title{Designing Machine Learning Toolboxes: \\ Concepts, Principles and Patterns}

\author{%
Franz J.~Kir\'{a}ly
\thanks{Corresponding author: \url{f.kiraly@ucl.ac.uk}}
\\UCL
\And
Markus Löning
\\UCL
\And
Anthony Blaom
\\University of Auckland
\And
Ahmed Guecioueur
\\INSEAD
\And
Raphael Sonabend
\\ UCL
}

\begin{document}
\maketitle

\begin{abstract}
Machine learning (ML) and AI toolboxes such as \software{scikit-learn} or \software{Weka} are workhorses of contemporary data scientific practice -- their central role being enabled by usable yet powerful designs that allow to easily specify, train and validate complex modeling pipelines. However, despite their universal success, the key design principles in their construction have never been fully analyzed. In this paper, we attempt to provide an overview of key patterns in the design of AI modeling toolboxes, taking inspiration, in equal parts, from the field of software engineering, implementation patterns found in contemporary toolboxes, and our own experience from developing ML toolboxes. In particular, we develop a conceptual model for the AI/ML domain, with a new type system, called scientific types, at its core. Scientific types capture the scientific meaning of common elements in ML workflows based on the set of operations that we usually perform with them (i.e.\ their interface) and their statistical properties. From our conceptual analysis, we derive a set of design principles and patterns. We illustrate that our analysis can not only explain the design of existing toolboxes, but also guide the development of new ones. We intend our contribution to be a state-of-art reference for future toolbox engineers, a summary of best practices, a collection of ML design patterns which may become useful for future research, and, potentially, the first steps towards a higher-level programming paradigm for constructing AI.
\end{abstract}

\section{Introduction}
\label{sec:intro}
Machine learning (ML) applications typically involve a number of steps: one first specifies, trains and selects an appropriate model and then validates and deploys it. ML practitioners write code to evaluate such workflows. Application code often incorporates classes and functions from one or more software packages called toolboxes. Toolboxes provide prefabricated pieces of code that make it faster to write software. Instead of constructing every piece of software from scratch, one can simply put together prefabricated pieces of code. Popular examples include \software{scikit-learn} \cite{Pedregosa2001} in Python, \software{Weka} \cite{Hall2009} in Java, \software{MLJ} \cite{Blaom2020a} in Julia, and \software{mlr3} \cite{lang2019mlr3} or \software{caret} \cite{Kuhn2008} in R. Toolboxes have become the backbone of modern data science and principal tool for ML practitioners in recent years.

The applicability and effectiveness of a toolbox depends crucially on its design. The design determines not only how easy it is for practitioners to understand and use toolboxes but also how easily they can be tested, maintained and extended. Despite the importance of toolboxes, the key design principles in their construction have never been fully discussed. In this paper, we try to close this gap by analyzing common designs of toolboxes and deriving reusable principles and patterns.

Much of the existing software literature on ML/AI has focused on best practices, design patterns and architecture for ML systems \cite{nalchigar2019solution, sculley2015hidden, nascimento2020software}, but little attention has been paid to design at the toolbox level, even though toolboxes are a central component in ML systems (i.e.\ kernel software) implementing core algorithmic and mathematical functionality. In fact, authors often point out that there is a lack of strong abstractions to support ML software, hindering much needed progress towards a more declarative SQL-like language for ML/AI \cite{sculley2015hidden, Zheng2014}. Similarly, in the ML literature, while the importance of toolbox development has been widely recognized \cite{Sonnenburg2007}, discussions on design are still rare. For example, toolbox developers often subscribe to a set of principles when presenting their work, however these principles typically remain abstract, with literature primarily discussing features or usage, and not design (see e.g.\ \citet{Buitinck2013}). While the actual software tends to contain a wealth of design ideas, we are not aware of literature references where data science specific patterns, or the process of deriving concrete implementations from design principles, are described or studied.

In this paper, we make a first attempt at consolidating the science of ML toolbox design. Rather than describing a single toolbox, or analyzing ML/AI software at the system or platform level, this paper focuses specifically on generalizable design principles and patterns for ML toolboxes. We investigate ML toolbox design by developing a well-grounded conceptual model for the ML domain. At its core, we propose a simple but powerful idea called scientific types -- a new type system which captures the data scientific purpose of key objects in the ML domain (e.g.\ data objects, probability distributions or algorithms).
When applying scientific typing to common ML objects, we are able to derive clear principles and reusable patterns that can not only explain central aspects of existing toolboxes, but can also guide the development of new ones.

Toolbox design -- much like any software design -- is hard. One has to identify the relevant objects, abstract them at the right level of granularity, define class interfaces, and specify hierarchies and relationships among them. In comparison to classical software engineering, ML/AI toolbox development raises a number of additional questions: In which respects is the ML domain different from other domains? How can we find useful abstractions in the ML domain? For example, how can we identify useful and meaningful categories for ML algorithms? How do we define interfaces for different algorithms? Our aim is to motivate and explain the principles and patterns we propose in a way that other developers, practitioners and decision makers who rely on ML toolboxes can refer to them when assessing designs.

A natural source of relevant ideas, useful formalism and best practices is found in the corpus of software engineering. Our approach largely falls under domain-driven design as put forward by \citet{Evans2004}, but also draws on ideas from ``design by contract'' \cite{Meyer1997}, ``responsibility-driven design'' \cite{Wirfs-Brock1989, Wirfs-Brock1990, Wirfs-Brock2003} and pattern-oriented design \cite{Gamma2002, Buschmann2007}. While much is direct transfer from existing software engineering practice, there is a substantial aspect in which ML departs from other domains: algorithms, interfaces and workflows are closely intertwined with mathematical and statistical formalism -- to an extent that mathematical objects are not only at the methodological core, but a key element in representation, workflow specification and user interaction.

This situation poses unique challenges which we attempt to address in what follows by a combination of formal mathematical statistics, well-grounded design principles based on our new scientific type system, adapted state-of-art architectural patterns from widely used toolboxes, and our own experience of designing multiple ML toolboxes in different languages (e.g.\ \software{sktime} \cite{Loning2019}, \software{MLJ} \cite{Blaom2020a}, \software{mlr3proba} \cite{Sonabend2020}, \software{pysf} \cite{Guecioueur2018}, \software{skpro} \cite{Gressmann2018a}, \software{mlaut} \cite{Kazakov2019}). In our analysis, we rely on object-oriented programming (OOP), the predominant paradigm for most complex ML/AI projects. While OOP is often assumed without rationale, we instead motivate specifically why OOP is especially well suited for ML toolboxes. Our analysis also extends in theory to other programming paradigms.

We intend our contribution to be a state-of-art reference for future toolbox engineers, a summary of best practices, a collection of ML design patterns which may become useful for future research, and, potentially, the first steps towards higher-level declarative programming languages for constructing AI.

\subsection*{Summary of contributions}
Our main contributions are both theoretical and practical:
\begin{itemize}
	\item As the basis for ML/AI toolbox design, we propose a theoretical or \emph{conceptual model for key objects in the ML/AI domain}. At its core, we develop a new type system called \emph{scientific typing} which captures the data scientific meaning of key ML objects (e.g.\ data objects, distributions or algorithms) and translates them into implementable software blueprints.
    \item From our analysis, we derive a set of practical, reusable \emph{design principles} and \emph{design patterns} that can motivate and explain existing toolbox designs and can guide future developments.
\end{itemize}

\subsection*{Structure}
The remainder of the paper is organized as follows: In section \ref{sec:problem_statement}, we define the problem we are trying to address in more detail. In section \ref{sec:cmodel}, we start developing our conceptual model for the ML/AI domain, which we then complete in section \ref{sec:cmodel.sci} by introducing our new scientific type system. From our conceptual model, we first derive a set of high-level design principles in section \ref{sec:principles}, and then a set of reusable design patterns in section \ref{sec:patterns}. In section \ref{sec:implementation}, we illustrate how our derived principles and patterns can not only explain existing designs, but also guide new ones. Section \ref{sec:conclusion} concludes.

\section{Problem statement}
\label{sec:problem_statement}

In this section, we state and contextualize the problem we are trying to address in more detail. It will be helpful to distinguish between two problems in ML research and applications:
\begin{description}[leftmargin=0pt]
	\item[Practitioner's problem.] The practitioner's problem is to solve a certain practical ML problem. For example, classifying emails as spam or forecasting the demand of electricity. This typically involves defining a suitable ML workflow from model specification to deployment. To run their workflows, ML practitioners write code specifically for the application at hand, but often incorporate functionality from existing toolboxes so that we do not have to write everything from scratch. We usually evaluate our solutions using a quantifiable loss function but also take into account other criteria such as interpretability, fairness or computational efficiency.
	\item[Developer's problem.] The developer's problem is to develop toolboxes that help practitioners solve their respective problem more effectively. For example, developing a toolbox for cross-sectional ML problems as done in \software{scikit-learn} \cite{Pedregosa2001}. A toolbox is a collection of related and reusable classes and functions designed to provide useful, general-purpose functionality. Developing a toolbox involves defining the scope or application domain we want to cover, identifying key objects, finding useful abstractions for them, and implementing them in a re-usable software package. We typically evaluate our toolbox in light of three aspects: its content (what the code does), design (its language and structure) and performance (how efficiently it runs). For example, we want to make sure that the code is correct, or consistent with some specification through testing, that it is readable, and that it is efficient enough to be practically useful in terms of run time and memory usage.
\end{description}

This paper is primarily concerned with the developer's problem, and more specifically with the design aspect. Design sometimes also refers to the design process. In this paper, we are mainly concerned with design as the result of that process, but, of course, the discussion in this paper may also inform the design process. Viewed as a result, toolbox design is about the overall structure and organization of software into components, but also the specifics of defining key components and mapping of mathematical concepts and operations onto classes and functions. Also note that design is not independent of the content and performance concerns. For example, more readable code may be less efficient but be easier to test. In general, there will be no ``pareto optimum'' and trade-offs have to be made.

Toolbox design is crucial to their applicability and effectiveness. On a high level, we believe that the better design choices for ML/AI toolboxes:
\begin{itemize}
	\item align the language and structure of code with ``natural'' mathematical, statistical and methodological semantics (see domain-driven design \cite{Evans2004}),
	\item facilitate user interaction in common data scientific workflows,
	\item facilitate rapid prototyping and scaling of AI pipelines,
	\item facilitate reproducibility and transparency of workflows and algorithms,
	\item facilitate checkability and validability of AI (e.g.\ with respect to algorithmic performance),
	\item are architecturally parsimonious, i.e.\ avoid unnecessary conceptualization, formalization and semantics,
	\item are inter-operable with other major tools in the same domain,
	\item facilitate maintenance and extensions.
\end{itemize}

While the importance of toolbox design has been recognized \cite{Sonnenburg2007}, there is still little research on ML/AI toolbox design. One reason may be that toolbox design seems obvious or trivial, especially in hindsight. After all, popular toolboxes have been so widely adopted that it is hard to imagine what better (or worse) alternatives would look like. However, it becomes clear that many questions remain unanswered when one tries to explain their design choices or when one tries to develop new toolboxes. Example \ref{code:sklearn} shows a typical supervised classification workflow as provided by \software{scikit-learn}.

\begin{minipage}{\textwidth}
\small
\begin{lstlisting}[
caption=Typical supervised classification workflow with \software{scikit-learn} in Python,
label=code:sklearn
]
classifier = RandomForestClassifier()
classifier.fit(y_train, X_train)
y_pred = classifier.predict(X_test)
\end{lstlisting}
\footnotesize{{\ttfamily X\_train} and {\ttfamily y\_train} denote the training data feature matrix and label vector, {\ttfamily X\_test} is the feature matrix of the test set, and {\ttfamily y\_pred} the predicted label.}
\end{minipage}\hfill

On a practical level, one may ask: why is ``classifier'' as in \texttt{RandomForestClassifier} a useful and meaningful algorithm category or type? Is there a formal, mathematical definition for such types? Why are the main interface points \texttt{fit} and \texttt{predict}? How is a ``classifier'' related to other algorithm types? How can they interact? On a methodological level, it is not even clear what our answers would look like: How do we identify, describe and motivate abstractions in the ML domain? How do we make sure that our abstractions are both closely linked to the underlying mathematical concepts but also easily translatable into software?

We believe that toolbox design remains, partly at least, an art, however that there exist some principles that can explain and guide our answers to these questions.
Throughout the paper, we make qualitative arguments to support our principles drawing on the adherence to common domain-driven design principles \cite{Evans2004}, design patterns \cite{Gamma2002, Larman2012} and other best practices from software engineering \cite{Pressman2005, Bass2003}.
Through our analysis, we believe that we cannot only explain why certain designs are more successful than others, but also gain new insights to improve future toolbox design.

\section{A conceptual model for AI frameworks}
\label{sec:cmodel}

In line with domain-driven design \cite{Evans2004}, we begin by outlining a conceptual model for concepts of relevance in the context of ML/AI software frameworks. In domain-driven design, development of a ``conceptual model'' is usually the initial step in which relevant concepts are delineated and defined. The software design is then derived to closely match the developed conceptual model. In this work, we go one step further and derive foundational design patterns for the AI framework setting -- that is, not just specific toolbox designs, but general guiding principles on how to design toolboxes. It is important to note that in this conceptualization, two perspectives are inextricably linked:
\begin{itemize}
	\item The realm of implementations: machine data types, algorithmics, input/output specifications,
	\item The realm of formal mathematics: formal domains, types, signatures, assumed mathematical object properties.
\end{itemize}
In consequence, our conceptual model fundamentally relies on ensuring that both perspectives are taken into account simultaneously.

In this section, we lay much of the conceptual groundwork which we will leverage in the next section to introduce scientific typing. As is standard in software engineering, we observe that in our domain of interest some objects vary more frequently than others. We start by identifying and separating out the more change-prone objects as sensible points for abstraction. These are:
\begin{itemize}
    \item Problems or tasks that we want to solve (e.g.\ regression or classification),
    \item Algorithms that can solve these tasks (e.g.\ linear regression or a decision tree),
    \item Related mathematical objects, on the formal level, such as loss functions or distributions.
\end{itemize}
In particular, we begin developing our conceptual model for AI frameworks by defining:
\begin{itemize}
    \item The \emph{interface viewpoint}, which introduces the dual algorithmic/mathematical perspective,
    \item The \emph{problem specification}: solving a formal ``learning task'' by adopting the dual perspective,
    \item The \emph{conceptual object model}: conceptualization of mathematical objects and algorithms in terms of mathematical and algorithmic formalism.
\end{itemize}

\subsection{What is ML from an interface perspective? -- the learning machine}
In practice, an ``AI'', ``learning machine'', or short ``learner'', is a collection of individual algorithms.
As a key step in our conceptualization, we adopt an interface viewpoint for learners -- that is, we consider a learning machine as defined by in which respect it interacts with its environment, e.g.\ other algorithms, sensors, a human data scientist.
\begin{figure}[tbh]
\centering
\includegraphics[width = 1.0\textwidth]{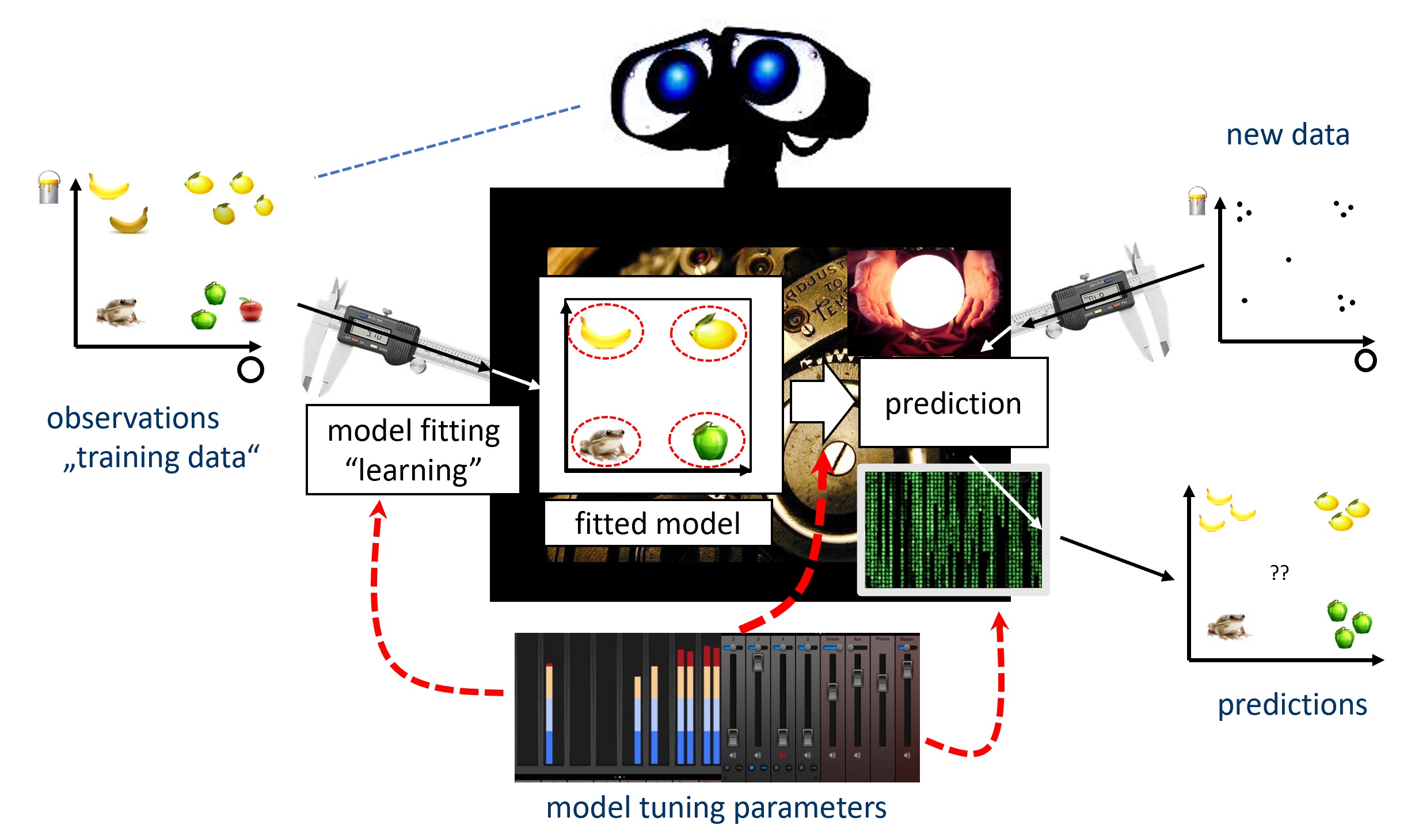}
\caption{Conceptual model for a supervised classification ``learning machine'', schematically depicted from the interface perspective. The ``learning machine'' is depicted as a ``black box robot'', with black arrows depicting data related interface points, and hyper-parameters (bottom box and red arrows) constituting another interface point. The figure also schematically depicts the sequence by which interface points appear in the most common user journey of the learning machine. From left to right: first (leftmost black arrow), the supervised learner ingests (labelled) training data to ``fit a model'' that explains the relation between x/y coordinates in the data (roundness, yellowness) and label (banana, lemon, frog, apple), from training data that consists of pairs of x/y coordinates and labels. Through the fitting interface, ``fitted model'' becomes part of the learning machine's internal state, depicted as x/y coordinate areas representing certain labels (red circles with label). This fitted model can then (rightmost black arrows) be used in prediction, for inferring labels for x/y coordinate points without a label. All operations may be influenced by hyper-parameter settings (box in the bottom, red dotted arrows depict this relation). The implementation of the interface points is not specified and can differ between different types of algorithms, e.g.\ random forest or k-nearest neighbors classifier -- but all classification algorithms follow this interface. The conceptual model for the supervised classification ``learning machine'' is the totality of this diagram, rather than any specific part of it -- that is, not only the model or its mathematical specification, but the entire interface with abstract specification of relations between inputs, outputs, model specification, hyper-parameter settings, etc.}
\label{figure:robot}
\end{figure}
Figure \ref{figure:robot} schematically depicts a supervised learner, our running example for much of this paper. Typically, in common data scientific practice, a user would interact with the supervised learner in one of the following ways:
\begin{itemize}
\item Specifying the model class, model parameters, or model structure,
\item Ingesting data (more precisely: feature-label pairs) to ``fit'' or ``learn'' the (statistical) model,
\item Calling a fitted model for prediction on new data (more precisely: features) to ``predict'' labels.
\end{itemize}
There are a number of other possible user interactions or interface points. For example, setting or retrieving hyper-parameters or fitted parameters, updating a fitted model in an online learning setting, or persisting a fitted model for later deployment.

Ultimately, the interface standpoint enables the ML practitioner to clearly specify what the algorithm does, as opposed to how it does ``it'' (or something that is not further specified). Conversely, systematic adoption of a clean interface design forces the practitioner -- as a measure of scientific hygiene and transparency -- to always specify the purpose that the algorithm is meant to solve, thus leading to higher quality code, in terms of robustness, transparency and readability.

We would like to make a number of additional remarks about the interface viewpoint:
\begin{itemize}
\item An interface specification can be used to define a \emph{type} of learner: whenever another learner \emph{has} the ``same'' interface, it is considered as \emph{having the same type}. For example, all supervised learners allow the modes of access described above -- specification, fitting, prediction -- and this can be used as a \emph{defining property} subject to assumptions on their statistical nature.
\item It is important to separate the concepts of the ``learner'' and possible \emph{workflows} it is part of. For example, predicted labels from a supervised learner can then be used to obtain performance certificates for the model. However, note that this is no longer an interaction directly with the learner, but with data artifacts which it has produced, namely the predictions. Conceptual separation of use cases, workflows, interfaces and methods is common in software engineering practice, but not always present (or necessary) in methodology-oriented data science. Conceptual hygiene in this respect allows a clear treatment of technological units, as well as clear formal separation of means from ends.
\item The interface viewpoint is a substantially different from common exposition in ML or data science, where learners are often defined in terms of a mathematical/statistical model, or the algorithms. However, this is not a contradiction, but complementary: both the mathematical and algorithmic specification, as well as the interface are important and necessary to fully specify a particular learner. In Figure \ref{figure:robot}, one may imagine the former ``inside'' and the latter ``outside'' of the ``black box robot''. The distinction between internal specification and interface allows to formally distinguish specific learners and types of learners, also a recurring theme in this paper.
\end{itemize}

For example, implementations of different supervised prediction strategies in common toolboxes such as \texttt{scikit-learn} share the same class/methods interface (e.g.\ \texttt{RandomForestClassifier} and \texttt{KNNClassifier}). This interface also abstracted from data representation in the sense that for application to data, only the reference to the supervised prediction strategy needs to be exchanged in the workflow. We will see how this conceptual property maps onto specific design patterns in section \ref{sec:patterns.ML}.

\subsection{No solution without a problem -- task specification}
\label{sec:cmodel.task}

A crucial aspect of the aforementioned necessity to address the mathematical set-up in architectural designs is a clean problem specification. Generally, ML software does not exist on its own in a vacuum, but serves a specific purpose. On the scientific side, statement of the purpose of an algorithm is necessary for validability -- if it isn't clear what an algorithm does, no argument can be made that it was better than doing nothing. On the practical side, only an architecture that allows clean problem specification is \emph{operationalizable} for reproducible and transparent data science (since absence of such architecture leaves goals unspecified and uncheckable).

We illustrate what this concretely means in the context of our running example, supervised learning. For this, we present a (simplified) ``consensus'' formulation of the (supervised) learning task, as the \emph{subject} of further discussion on architecture (i.e.\ an example of the practitioner's problem).\footnote{For a more detailed overview of supervised learning, see e.g.\ \citet{Hastie2009}.}

\begin{tcolorbox}[title=The supervised learning task]
\begin{description}
\item[Data specification.] Consider data consisting of feature-label pairs $(X_1,Y_1),\dots, (X_N,Y_N)\simiid (X,Y),$ where $(X,Y)$ is a generative random variable taking values in $\calX\times \calY$. The domain $\calX$ is a finite Cartesian product of primitive data types (real, finite/categorical, strings). The domain $\calY$ is either continuous, i.e.\ $\calY\subseteq \RR$, or of finite cardinality, i.e.\ $\card\calY < \infty$. In the first case, the task is called ``supervised regression'', in the second ``supervised classification''.
\item[Definition of learning.] The goal is to construct a supervised learning functional $f$, taking values in $[\calX\rightarrow \calY]$. The functional $f$ may depend on the \emph{values} of the feature-label pairs $(X_1,Y_1),\dots, (X_N,Y_N)$, but does not have access to the (distribution) \emph{law} of $(X,Y)$.
\item[Definition of success.] A (non-random), \emph{fixed functional} $g:\calX\rightarrow \calY$ is considered performant/good if the expected generalization loss $\EE[L(g(X),Y)]$ is low for some user specified loss function $L:\calY\times \calY\rightarrow \RR$, e.g.\ the squared loss $L:(\widehat{y},y)\mapsto (\widehat{y}-y)^2$ for regression, or the misclassification loss $L:(\widehat{y},y)\mapsto \OOne [\widehat{y}\neq y]$ for classification.
A \emph{learning algorithm} $f$, possibly statistically dependent on $(X_1,Y_1),\dots, (X_N,Y_N)$, is considered performant/good if the expected generalization loss $\EE[L(f(X^*),Y^*)]$ is low, where
$(X^*,Y^*)\sim (X,Y)$ is independent of $(X_1,Y_1),\dots, (X_N,Y_N)$ and $f$.
\end{description}
\end{tcolorbox}

The three parts of the learning task are crucial. For a clear problem specification, we need to know: given what, what are we doing, and how do we know we did it (well)? More precisely, a task description contains:
\begin{enumerate}
\item[(a)] a formal {\bf data structure specification}. This must include the \emph{relational structure} (in database parlance) and \emph{statistical structure} assumptions (usually generative). In the supervised learning example, the relational structure is that of a single data table, rows being index $i$ of $X_i,Y_i$, columns being variables/components of $X_i$, with one column designated as the prediction target ($Y$). The statistical assumption is simple but crucial: rows are assumed to form an i.i.d.~sample.
\item[(b)] a {\bf learning interface specification}. That is, at what point does the learner ingest what information and what does it return? This is a combination of an \emph{algorithmic interface}, what objects are related as inputs and outputs of subroutines, and a \emph{statistical interface}, in specifying the statistical dependence structure. In the case of supervised learning, the functional $f$ is \emph{computed} from the data $X_i,Y_i$. As a mathematical object, $f$ in general depends on the $X_i,Y_i$, and is statistically independent of anything else outside the learner. The specific algorithm by which $f$ is computed is unspecified, but needs to follow the given form for the task.
\item[(c)] a {\bf success specification}. That is, how is success defined? For a workable architecture, this needs to be \emph{operationalizable} -- i.e.\ whether the goal has been reached should be clearly defined, empirically checkable in the scientific sense, and algorithmically defined, subject to valid reasoning. Usually, the goal specification is mathematical-statistical, and arguing that a goal is sensible relies on theoretical arguments. In the supervised learning example, the goal is the aim of achieving a low generalization loss, and it may be measured by algorithms for quantitative performance estimation and comparisons of algorithms.
\end{enumerate}

We make a number of additional remarks:
\begin{itemize}
\item Our conceptual choice to insist on a goal specification together with the learner specification may seem unusual. We argue that an AI needs to follow some checkable goal, otherwise its usefulness is not only dubious, but in fact entirely unfalsifiable in the sense of the scientific method. Thus, not stating the goal in using a learner is unscientific or pseudo-scientific. Note that in many toolboxes like \software{scikit-learn} or Weka, the task is already implicitly specified by the choice of algorithm type and scope of the toolbox (e.g.\ when we choose a regressor from scikit-learn, the implied task is tabular regression).
\item Of course, the goal specification does not need to involve prediction of something, or predictive modelling in general. Model parameter or structure inference (e.g.\ inspection of variable coefficients) in a context of a model interpretability task, or causal modelling (e.g.\ building decision rules) in the context of a decision making task are common, important goals which are not strictly predictive, but may happen in the context of predictive models. In either case, there is still a need to state \emph{operationalizable} criteria for what makes algorithmic outputs sensible, according to the previous argument.
\item A task specification may be seen to specify interfaces through its change-prone parts: in the supervised learning example, the data and its type (how many columns etc); the specific choice of mathematical and algorithmic form for the learner $f$; the specific choice of loss function $L$. These objects are interchangeable, as long as they comply with the task specification. From an architectural perspective, they may be seen as natural interfaces, and conceptual fixed points to leverage for interface design.
\item As a subtle point on success specification, there is a difference between operationalizing criteria for what makes a sensible success control (or evaluation) procedure, and the success control procedure itself. In the supervised learning example, the expected generalization loss is an unknown (generative) quantity, therefore additional algorithms are needed for its estimation. These algorithms may take different forms, and may be sensible to varying degree, according to mathematical argument or proof. An operationalizable goal specification can be leveraged to check whether such algorithms claim to conduct a suitable form of success control.
\end{itemize}

We proceed by illustrating the necessity of items in our task model in a less common example: the learning task of forecasting. Under the name of ``forecasting'', related but distinct tasks are often conflated in literature.\footnote{For an introduction to forecasting, see e.g.\ \citet{Box2013}.} One reason is that the description of the problem commonly contains only the data structure specification, but not a learning interface specification. This situation may lead to pairing the wrong strategy with the problem at hand, or comparing pears with apples. More concretely, a consensus data specification (for the purpose of simplified exposition, equidistantly-spaced stationary data) plus a common specification of learning, as follows:

\begin{tcolorbox}[title=The forecasting task]
\begin{description}
	\item[Data specification.] The data is a partial observation of a stationary sequence of random variables $(Z_i,i\in \NN)$ taking values in $\mathcal{Z}\subseteq\RR$; at a given point in time $\tau$, only $Z_{\le \tau} := (Z_i, i\le \tau)$ are observed.
	\item[Definition of learning.] The goal is to produce forecasts $\widehat{Z}_j$ for $Z_j,$ for certain $j$.
\end{description}
\end{tcolorbox}

The problem with this task specification (apart from the obvious lack of success control) is the incomplete algorithmic and statistical learning interface definition: what mathematical and algorithmic form does a learner take? What may $\widehat{X}_j$ depend on, algorithmically and statistically? There are, in fact, (at least) two very common task definitions this could be completed to:

\begin{tcolorbox}
\begin{description}
	\item[Definition of learning, variant 1 (fixed horizon).] The goal is to produce forecasts $\widehat{Z}_j$ for $Z_j,$ for $j= \tau+1,\dots, t$. For this, only $Z_1,\dots, Z_\tau$ are available as algorithmic input to the learner.
	\item[Definition of learning, variant 2 (sliding window).] The goal is to produce forecasts $\widehat{Z}_j$ for $Z_j,$ for $j= \tau+1,\dots, t$. To produce $\widehat{Z}_j$, the learner has algorithmic access to, and only to, $\widehat{Z}_{j'}$ with $j'\lneq j$. In particular, $\widehat{Z}_j$ must be statistically independent of $Z_{\ge j}:=(Z_i, i \ge j),$ conditional on $Z_{<j}:=(Z_i,i\lneq j)$.
\end{description}
Suitable goal specifications also diverge accordingly.
\end{tcolorbox}

In summary: without a clean task model that separates the two cases, an interface could have conflated ``forecasters'' which are statistically incompatible. Worse, it could have done so under an identical data model and algorithm interface -- not entirely dissimilar to the problematic situation where there are multiple instances of the same power outlet, but providing electricity at different voltages or currents\footnote{i.e.\ the benefit of being able to plug in all appliances everywhere is offset by the risk of their accidental destruction, making this interface design choice somewhat questionable, as long as accidental destruction is an undesired feature.}. Thus, without accounting cleanly for both algorithmic and statistical-mathematical aspects, a satisfactory AI interface architecture cannot be constructed. 

\subsection{The conceptual object model}
\label{sec:cmodel.objects}

Until now, we have discussed the importance of clearly specifying learning problems and the interface viewpoint for algorithms. We proceed by conceptualizing key objects found in typical ML workflows, including algorithms among others.

There are a number of (conceptual) objects that appear repeatedly as variable parts or interface parts in a ML framework. On the conceptual level, these objects are all of formal mathematical and algorithmic nature. Examples of recurring objects are:
\begin{itemize}
\item Data frames with definable and inspectable column types, e.g.\ numerical, categorical, or complex (in a formal mathematical sense),
\item Non-primitive and composite data objects such as series, bags, shapes, or spatial data,
\item Sets and domains, e.g.\ for specifying ranges, intervals, parameter domains,
\item Probability distributions, e.g.\ arising as return types in Bayesian inference or probabilistic prediction,
\item ML algorithms (sometimes also called estimators, strategies, models or learning machines),
\item ML pipelines, networks, composites or entire workflows,
\item Workflow elements such as for automated tuning or model performance evaluation.
\end{itemize}

An important distinction that divides these objects into two categories is that of \emph{statefulness}. Being an entity object (also: immutable) or a value object (also: mutable) is a common distinction in computer science:
\begin{itemize}

\item \emph{Value objects} have the same content and properties independent of the condition or context
\item \emph{Entity objects} object have a continuous identity, and may be subject to change of content or properties
\end{itemize}

The typical ``mathematical object'' from classical mathematical notation is a value object -- an object that, once defined in common mathematical formalism (e.g.\ ``let $x:=42$''), remains unchanged within the scope of its existence.
Whereas ``learning'' objects, such as ML algorithms, are conceived of as entity objects -- in fact, their conceptualization is typically characterized by behavior at state changes, e.g.\ at ingestion of training data.

In our conceptual model we will hence single out \emph{mathematical objects} that model ``formal mathematical objects'' in the classical sense and are value objects, and \emph{learner objects} that model learning machines and strategies, and which are \emph{entity objects}. An ambiguous case to which we will return later is the nature of complex specifications, e.g.\ of algorithms or workflows, which are themselves value objects, but which specify entity objects.

In the list of common example above, the following are mathematical objects in this sense:
\begin{itemize}
\item Data frames with definable and inspectable column types, e.g.\ numerical, categorical, or complex,
\item Non-primitive and composite data objects such as series, bags, shapes, or spatial data,
\item Sets and domains,
\item Probability distributions.
\end{itemize}

Note that we consider data frames and data objects as value objects, because state change is not a defining conceptual characteristic of data (although state change may be defining for the special case of data streams). Change of the raw data or data cleaning -- as opposed to algorithmic feature extraction -- is something we do not conceptualize as ``within'' a ML framework. Overall, value objects are also those for which there is typically unambiguous convention on mathematical notation in the ML literature.

On the other hand, the following are learner objects, and thus entity objects:
\begin{itemize}
\item ML algorithms,
\item ML pipelines, networks, composites, or entire workflows,
\item Workflow elements such as for automated tuning or model performance evaluation.
\end{itemize}
All of these are defined by operations or actions which change an intrinsic ``state'' of object. For example, in ML strategies, typically one or multiple data ingestion steps are carried out, e.g.\ ``fitting'', which changes the strategy's state from ``unfitted'' to ``fitted''.

It is important to note that the ML literature disagrees on conceptualization of ML strategies in terms of mathematical formalism already: for example, in the supervised learning literature, an algorithm may be represented by as a prediction functional on test data, or a functional in which both training and test data have to be substituted, thus representing the fitting process by some of its argument (compare, for example, the framing in \citet[][chap. 2]{Hastie2009} and \citet[][chap. 1]{Bishop2006}). Besides the lack of consistency, we argue that there is also a lack of expressivity which prevents clear conceptualization on both mathematical and algorithmic levels. We will attempt to remedy this situation partly by introducing some formalism.

We continue with further details on the conceptual models for value and entity objects.

\subsection{The conceptual model for mathematical objects}
\label{sec:cmodel.mathobj}

Typically, mathematical objects (as previously introduced) appear both as \emph{internal concepts}, e.g.\ in representation or on internal interfaces, as well as elements of \emph{user interaction}, e.g.\ in interacting with data, specifying models, or inspecting results of modeling.

In either case, there are common interface and interaction cases in dealing with mathematical objects, i.e.\ what one would informally call a ``formal mathematical object''. These common interface/interaction cases are:
\begin{itemize}
\item[(i)] {\bf Definition:} creating inspectable variables that are mathematical objects. For example: ``let $X$ be a normal random variable with mean $0$ and variance $42$; or, ``let \texttt{myseries} be a discrete time series with values in $\RR^3$, of length $240$, time stamps being [...] and values being [...]''.
\item[(ii)] {\bf Value or parameter inspection:} ``what is the mean of $X$''; or, ``what is the value of \texttt{myseries} in January 2020?''
\item[(iii)] {\bf Type inspection:} querying the mathematical type (e.g.\ domain) of an object: ``what domain does $X$ take values in?''; or, ``how long is \texttt{myseries}?''
\item[(iv)] {\bf Inspection of traits and properties:} ``is the distribution of $X$ symmetric?''; or, ``are the time stamps of \texttt{myseries} equally spaced?''
\end{itemize}

By definition, state change is not an interface case for value objects, since the definition characterizes value objects as objects without such an interface or interaction point. Qualitatively, these are also not properties one would commonly associate with mathematical objects, as per implicit conceptualization and common use. Thus, ``classical'' mathematical objects are always value objects in our conceptual model.

Based on these interface cases, we postulate that the following are intrinsic attributes of a mathematical object:
\begin{itemize}
\item[(a)] a {\bf name or symbol}, e.g.\ $X$, or \texttt{myseries}
\item[(b)] a {\bf type}, e.g.\ a ``real random variable'', or ``real univariate time series''
\item[(c)] a {\bf domain}, e.g.\ ``the set $\RR$'', or ``absolutely continuous distributions over $[0,1]$''
\item[(d)] a {\bf collection of parameters} on which the object may depend, being mathematical objects themselves, in particular possessing name, type and domain themselves\footnote{Of course, there may not be infinite recursion. Therefore, eventually there are nested parameters without parameters.}
\item[(e)] {\bf traits and properties}, e.g.\ ``this random variable has a finite domain'', or ``this time series is univariate''
\end{itemize}

We postulate these attributes as inextricable attributes based on the related interaction and interface cases. In particular, we postulate that every mathematical object should be considered with a domain (which could be the universal domain), and a designated collection of parameters and properties (which could be the empty collection).

As an example for a mathematical object and its attributes, consider the common definition ``Let $X$ be a r.v., distributed according to the normal $\calN (0,42)$''. Here, one could define (a) type being ``random variable''; (b) being ``real random variable'', (c) domain being ``absolutely continuous r.v.~over $\RR$; (d) parameters being $\mu:= 0, \sigma^2:=42$, with types of $\mu,\sigma$ numeric/continuous, domains being $\mu\in \RR, \sigma^2\in \RR^+$; and (e) example traits as $X$ being symmetric, sub-Gaussian, mesokurtic.

An important subtlety here is defining a mathematical object with variable free parameters, in contrast to a mathematical object with variable but set parameters, e.g.\ a generic $\calN (\mu, \sigma^2)$ in contrast to a concrete $\calN (0,42)$.


In terms of implementation:
Attributes (a) and (b) are already commonly found as defining attributes of any variables in the base language of common programming languages (e.g.\ in Python or R usage) and can therefore be regarded as standard even beyond conceptualization of mathematical objects. Attributes (c), (d) and (e) are typically not found in vanilla expressivity and need dedicated implementation beyond the base language.
This will be addressed by design patterns in section \ref{sec:principles}. We will return to implementation details after completing the outline our conceptual model.

\subsection{Prior art}
Many aspects of our conceptual model are commonly found in the framing of ML theory, or software toolboxes -- however, to our knowledge, this has not been stated as a full conceptual model, or in the context of software design. The concept of a ``learning task'' is also not uncommon in taxonomies of data scientific problems, but typically implicit -- it is rarely made explicit as a concept on its own, possibly since it is not a necessary concept in typical methodological arguments where it appears as part of the framing. As a partial inspiration of our conceptual model, the \software{mlr} toolbox ecosystem \cite{Bischl2016, lang2019mlr3}  and \software{openML} API \cite{Vanschoren2014} has software formalism for (concrete) learning tasks, which is similar to our (generic) tasks on the conceptual level.

\section{Scientific types -- a conceptual model for formal objects}
\label{sec:cmodel.sci}

The full conceptual object model that we propose relies on the new idea of ``scientific typing'' which we introduce in this section.
A scientific type enable us to abstract mathematical objects, based on the set of operations that we usually perform with them (i.e.\ their interface) and key statistical properties, in a way that is both precise enough to be usable in mathematical statements, but also implementable in software. For example, we may say that a learning algorithm is of the type ``forecaster'' where the type in question has a precise mathematical meaning and an associated software template. To introduce this idea, we extend classical mathematical formalism to stateful objects, along the lines of von Neumann formalism and modern compiler theory \cite{Pierce2002}. 

In the following sections, we will leverage the idea of scientific typing to derive software design principles and patterns for implementing the respective formal mathematical objects (e.g.\ algorithms) in software.

\subsection{Scientific typing for mathematical objects}
\label{sec:cmodel.mathobj-sci}

Having established conceptualization of mathematical objects, we now introduce some accompanying mathematical formalism.
For this, we make use of common formalism in type theory, extended with some less common and novel aspects.
As we would like to make parameters and properties explicit attributes, we leverage and extend classical notation and type theory.

The general concept of a ``type'' arises both in mathematical formalism, as well as in modern programming languages \cite{Pierce2002}. In both frameworks, any object may have one or multiple of the following:
\begin{itemize}
\item[(a)] a {\bf name or symbol}, e.g.\ $x$
\item[(b)] a {\bf formal, symbolic type}, e.g.\ an ``integer''
\item[(c)] a {\bf value}, e.g.\ 42
\end{itemize}

In mathematical notation, the ``typing colon'' is used to denote an object's type, e.g.\ the formal statement $x :$ \texttt{int} for stating that the object denoted by symbol $x$ is of type \texttt{int}. Seasoned programmers will be familiar with the notation in their favored typed language\footnote{Even in weakly typed, duck typed, or untyped languages, there are usually mechanisms at the compiler or interpreter level that correspond to formal typing, even if they are not exposed to the user directly.}, which usually depends on the language (for example, the statement \texttt{int x} in Java).

Assigning a type to a symbol or value is a purely formal act and in-principle arbitrary; it becomes practically useful through typing rules (and/or mathematical axioms). Most importantly, any type has \emph{inhabitants}, that is, values that are admissible for a type. For example, the formal type ``integer'' is inhabited by the numbers 0, 1, -1, 2, -2, and so on. As such, mathematical types can often be identified with their inhabitant sets\footnote{Footnote for readers with a theoretical mathematical or methodological background: It should be noted that the statement $x:\ZZ$ (``$x$ is of type $\ZZ$) is different in quality from saying $x\in \ZZ$ (``$x$ is an element of the set $\ZZ$''), even if, as a statement, the one is true whenever the other is. The difference lies in that the first is interpreted as an \emph{attribute} of the \emph{object} represented by $x$ (``the object represented by the symbol $x$ has integer type''), the second as a \emph{property} of the \emph{value} of $x$ (``the value of the object represented by $x$ is an integer''). While this may not be much of a difference on the \emph{conceptual level} in mathematics, where symbols only serve as mental reference, there is a substantial difference on the \emph{implementation level} in a computer, where the symbol may be attached to an object or in-memory representation which implements the conceptual model.
On the formal side, this interpretation is also reflected through differences in axiomatization of type systems, versus axiomatization of set theory.
}, such as stating $x:\ZZ$ for an integer symbol $x$.

From primitive types, type operators allow construction of more complex composite types. Examples are the conjunction\footnote{One also finds the symbol ``$\sqcap$'' in literature.} (which can be identified with the Cartesian product $\times$ for sets) and the arrow ``$\rightarrow$'', which allow to construct function types. This is developed in Table \ref{tab:basictt} by examples that should highlight the general pattern.

\begin{table}[tbh]
	\centering
	\small
	\caption{\label{tab:basictt} Some examples of type specification in pseudo-code and formal mathematical type notation}
\begin{tabularx}{\textwidth}{XXX}
	\toprule			
	Description & Pseudo-code (typed) & Mathematical (type theoretic) \\
	\midrule
	$x$ is an integer & \texttt{int x} & $ x : \ZZ$ \\ \midrule
	$x$ is the integer $42$ & \texttt{int x = 42} & $ x = 42 : \ZZ$ \\ \midrule
	$f$ is a function which takes as input an integer and outputs an integer & \texttt{func int f(int)} & $f : \ZZ \rightarrow \ZZ$ \\ \midrule
	$f$ is the function which takes as input an integer and outputs its square & \texttt{func nat f(int x) return x*x} & $f : \ZZ \rightarrow \NN$; $x\mapsto x^2$ \\ \midrule
	$x$ is a pair made of: an integer (1st), and a real number (2nd) & \texttt{[int,real] x} & $x : \ZZ\times \RR$ \\ \midrule
	$f$ is the function which takes as input two integers and outputs their sum & \texttt{func int f(int x, int y) return x+y} & $f : \ZZ\times \ZZ \rightarrow \ZZ$; $(x,y)\mapsto x+y$ \\
	\bottomrule
	\end{tabularx}
	\begin{minipage}{\textwidth}
	\emph{Notes}: We express the example statements in vernacular description (first column), pseudo-code of a stylized typed programming language (second column), and formal mathematical type notation with inhabitant-set identification convention (third column). Mathematical type operators used in the third column are the conjunction type operator, denoted as $\times$, and the arrow (type construction or function) operator, denoted as $\rightarrow$. A reader may recognize the colon commonly used for range/domain specification in mathematical function definition as the typing colon, and the arrow in function definition as the arrow operator (such as in third/fourth row, third column); whereas the typing colon for domain statement of a variable (such as in first row, third column) would be unusual in contemporary usage, outside of formal type theory.
	\end{minipage}
\end{table}

Another important composite type is that of a structured type. Concretely, a class type is a container with named types in ``slots'' (formally: component types) which correspond to (object) variables or functions. The inhabitants of a structured type are structured objects, where every slot is inhabited by concrete value inhabiting the slot type.  While a more formal discussion can be found in \cite{Pierce2002}, there are no commonly used notational conventions. We proceed with the following ad-hoc conventions, inspired by UML class diagrams \cite{Larman2012}:

\begin{itemize}
\item Structured types are denoted by a Python-like \texttt{struct type} header, followed by the slots in the composite, with name and type. Slots are symbol/type pairs, separated by a typing colon.
\item Inhabitants of structured types are called structured objects and denoted by a \texttt{struct} header, followed by slots in the composite, with name and value. Slots are symbol/value pairs, separated by a typing colon.
\item The slots are categorized under one of three headers: \texttt{params}, for ``parameter'' slots that are immutable per inhabitant (sometimes also called static); \texttt{state} for state variables; and \texttt{methods}, which are of function type and correspond to object methods.
\item Implementations (and behavior) of methods can always (but need not) depend on parameters in \texttt{params} as function parameters.
\item If $a$ is a structured type or structured object, with a slot of name $y$, the slot can be directly referred to by the symbol $a.y$, i.e.\ the symbol for the structured type/object, followed by a dot, followed by the symbol for the slot name.
\end{itemize}

For example, the structured type \texttt{GaussianDistributionSignature} which represents the (input/output signature type) of a Gaussian distribution object (represented by cdf and pdf) could be defined as follows:
\begin{table}[H]
\begin{tabular}{l l c l}
	\multicolumn{4}{l}{\texttt{struct type GaussianDistributionSignature}}\\
    \quad \texttt{params} & $\mu$  & : & $\RR$\\
    \quad & $\sigma$  & : & $\RR^+$\\
	\quad \texttt{methods} & \texttt{cdf} & : & $\RR\rightarrow [0,1]$ \\
	\quad & \texttt{pdf} & : &  $\RR \rightarrow [0,\infty)$  \\
\end{tabular}
\end{table}
An inhabitant of this type would be the structured object
\begin{table}[H]
\begin{tabular}{l l c l}
    $a:=$ &&&\\
	\multicolumn{4}{l}{\texttt{struct}}\\
    \quad \texttt{params} & $\mu$  & = & $1$\\
    \quad & $\sigma$  & = & $4$\\
	\quad \texttt{methods} & \texttt{cdf} & = & $x \mapsto \Phi \left(\frac{x-\mu}{\sigma}\right)$ \\
	\quad & \texttt{pdf} & = &  $x\mapsto \frac{1}{\sigma \sqrt{2 \pi}} \exp\left(-\frac{1}{2}\left(\frac{x-\mu}{\sigma}\right)^2\right)$  \\
\end{tabular}
\end{table}
where $\Phi$ is the standard normal cdf. As a typing statement, we can write $a: \texttt{GaussianDistribution}$. To refer to the cdf, we can write $a.\texttt{cdf}$ which is identical to the function $x \mapsto \Phi \left(\frac{x-1}{4}\right)$

It is important to note that the type \texttt{GaussianDistributionSignature} does not fully define what is a Gaussian distribution object (as represented by pdf/cdf). In computer scientific terms, the \emph{implementation} is missing; in mathematical terms, the \emph{specification} or \emph{defining property} is missing. For example, the structured object
\begin{table}[H]
\begin{tabular}{l l c l}
	\multicolumn{4}{l}{\texttt{struct}}\\
    \quad \texttt{params} & $\mu$  & = & $1$\\
    \quad & $\sigma$  & = & $4$\\
	\quad \texttt{methods} & \texttt{cdf} & = & $x \mapsto 2\mu$ \\
	\quad & \texttt{pdf} & = &  $x\mapsto 42$  \\
\end{tabular}
\end{table}
also formally inhabits the structured type $\texttt{GaussianDistributionSignature}$.

To fix this issue, one can now define the \emph{scientific type} of a Gaussian distribution object by adding the compatibility between the parameters and methods into the definition. For example, by defining the following: ``An object $d$ has type \texttt{GaussianDistribution} if it has structured type
\begin{table}[H]
\begin{tabular}{l l c l}
	\multicolumn{4}{l}{\texttt{struct type}}\\
    \quad \texttt{params} & $\mu$  & : & $\RR$\\
    \quad & $\sigma$  & : & $\RR^+$\\
	\quad \texttt{methods} & \texttt{cdf} & : & $\RR\rightarrow [0,1]$ \\
	\quad & \texttt{pdf} & : &  $\RR \rightarrow [0,\infty)$  \\
\end{tabular}
\end{table}
and if the methods have the following form: $d.\texttt{pdf}(x) = \Phi \left(\frac{x-d.\mu}{d.\sigma}\right)$, and $d.\texttt{cdf}(x) = \frac{1}{d.\sigma \sqrt{2 \pi}} \exp\left(-\frac{1}{2}\left(\frac{x-d.\mu}{d.\sigma}\right)^2\right).$''

While it is somewhat cumbersome in this case, we have defined, a \emph{scientific type} \texttt{GaussianDistribution}.
In our conceptual model, a scientific type combines:
\begin{itemize}
\item A structured type signature with parameters made explicit,
\item Assumptions on formal properties that inhabitants of the type need to satisfy, in the form of mathematical definition and extrinsic to the structured type signature.
\end{itemize}
In particular, any object with the type \texttt{GaussianDistribution} will, by definition, satisfy the compatibility between parameters and methods expected intuitively from a Gaussian distribution. We can also define inhabitant sets in a similar way to define domains.

Scientific types, or \emph{scitypes} for short, as introduced above by example, satisfy the postulates from our conceptual model of mathematical objects in the following way:
\begin{itemize}
\item[(a,b)] name and type statement are intrinsic to the notation/definition of the structured type
\item[(c)] domains can be specified as inhabitant sets
\item[(d)] parameter statements are intrinsic to our extended notation/definition of the structured type, in the \texttt{params} block.
\item[(e)] traits and properties are encoded by the slots in the \texttt{methods} block.
\end{itemize}

A reader may like to note a few things:
\begin{itemize}
\item While the initial definition of a scitype may be a bit cumbersome, later reference to the type or an arbitrary inhabitant can be invoked with a few symbols, e.g.\ ``let $d : \texttt{GaussianDistribution}$'', and methods or parameters may also be invoked without prior appearance in the binding statement, e.g.\ $d.\mu$.
\item The conceptualization resolves some items of common ambiguity of mathematical language. For example, which symbols are to be considered parameters? The resolution is: parameters are explicitly designated as a matter of definition. Or more subtle ambiguities around concept identity such as of a ``distribution'' which, across literature, may be defined or denoted in terms of different methods or properties (e.g.\ distribution defining functions, measures, maps, etc), depending on source and taste of the author. The resolution is: the object need not be made or defined identical with any of its methods, it ``has'' the methods as attributes.
\item Subject to some formalization burden, the concept of scientific types can be ``properly'' formalized, along the lines of classical type theory. This is not complex and simply amounts to defining inhabitance by satisfaction of a definition property, and should be intuitively clear from the example -- for readability, we refrain from adding further formal overhead.
\end{itemize}

Scitypes for mathematical object may not be directly necessary or fully novel in isolation. However,  in the next section, we will further build on scitypes for mathematical objects to complete the conceptual model for learner objects.

\subsection{Scientific typing -- towards a language of ML toolboxes}
\label{sec:cmodel.ML-sci}

We continue with introducing our conceptual model for learner objects (which are entity objects), along with an extension of the scitype concept introduced above, which will be crucial to finalize our proposed conceptual model.

A central property of modern AI methods and pipelines is their statefulness and compositionality -- that is, objects changing their state in interaction with data, and consisting of multiple parts whose algorithmic methods interact in intricate ways. Common examples are typical supervised pipelines ingesting data, extracting features, making predictions; or, neural networks, constructed from layers between which information is propagated. We will use these two examples -- pipelines and neural networks (in their simplest form) -- as running examples for the discussion in this section. The goal is to further develop our conceptual model by developing a formal answer to the question ``what is a pipeline'' or ``what is a neural network''.

In contemporary toolboxes such as \software{scikit-learn} (pipelines) and keras (neural networks), the architectural design mirrors the common mathematical or methodological conceptualization of composition. In the examples, this composition is of different types of estimators in pipelines (e.g.\ feature extractors or predictors) or layers in neural networks. The composition design is usually implemented through so-called structural object-oriented patterns, which we will discuss in greater detail later on. The goal for this section is to explain the general ideas by which we can make precise the above in the two running examples. We later show how these give rise to natural architectural patterns.

Ultimately, we would like to be able to make precise statements that go beyond vague and qualitative description, i.e.\ the formal counterpart of, say, an intuitive depiction such as in figure \ref{figure:robot}, within a formalism which allows:
\begin{itemize}
\item[(i)] Formal conceptualization of, and distinction between, different kinds of algorithms and construction principles,
\item[(ii)] Abstract definitions involving these, such as ``a sequential pipeline is a mapping $(\texttt{Transformer})^n\times \texttt{SupervisedLearner}\rightarrow \texttt{SupervisedLearner}$ with the following properties [...]'', or ``a multi-layer neural network is constructed as $\texttt{InputLayer}\times (\texttt{Layer})^n\times \texttt{OutputLayer}\rightarrow \texttt{NeuralNetwork}$'',
\item[(iii)] Formally meaningful expressions for specification of algorithmic composites such as \texttt{Pipeline(PCA(),SVM())} or \texttt{NeuralNetwork(InputLayer(), Layer1(), Layer2(), OutputLayer())}, similar to the syntax in state-of-art toolboxes (e.g.\ \software{scikit-learn}, \software{TensorFlow} \cite{Abadi2016} or \software{PyTorch} \cite{Paszke2019}).
\end{itemize}

We argue that the natural formal framework for a conceptual model of AI architectures is rooted in the theory of higher-order type systems, which are at the foundation of modern programming languages, as well as of modern mathematical formalism (often implicit). The reason is that higher-order type systems allow to make explicit the kind of component, its operations, their dependency on settable parameters and input-output relationships, in a precise way that is usable within mathematical statements, but also as a basis for computation-based reasoning that is machine implementable. As such, identifying the ``right'' higher-order type system for AI architecture is also the first step towards an AI-specific higher-level declarative language for model specification, execution, and deployment -- similar to how modern programming languages are removed in abstraction from assembly code specification.

Such a type system can be achieved by extending the scientific typing system for mathematical objects presented above -- resulting in a formal conceptual framework for discussing types of ML components.
\begin{itemize}
\item Class types are denoted by a Python-like \texttt{class type} header, followed by the slots in the composite, with name and type.
\item The slots are categorized under one of three headers: \texttt{params}, for ``parameter'' slots that are immutable per inhabitant (sometimes also called static); \texttt{state}, for mutable slots that change with state changes of an inhabitant; \texttt{methods}, which are of function type and correspond to object methods.
\item Implementations (and behavior) of methods can always depend on parameters in \texttt{params} as function parameters.
\item When specifying a type, a name of a symbol in scope can be used instead of its type. If a symbol in \texttt{state} is used in this way in a function in \texttt{method}, the function is interpreted to write to, or read the value from that state.
\item Inhabitants of a class type are defined by values for \texttt{params} and \texttt{methods}, while the \texttt{state} is not made explicit (as it can be changed by objects).
\end{itemize}

For example, the class type of a class \texttt{PowerComputer} which computes the $n$-th power of a number in one of its slots could be
\begin{table}[H]
\begin{tabular}{l l c l}
	\multicolumn{4}{l}{\texttt{class type PowerComputer}}\\
    \quad \texttt{params} & $n$  & : & $\NN$\\
	\quad \texttt{state} & $x$  & : & $\RR$\\
	\quad \texttt{methods} & \texttt{store\_number} & : & $\RR\rightarrow x$ \\
	\quad & \texttt{compute\_power} & : &  $x \rightarrow \RR$  \\
\end{tabular}
\end{table}
Here, the exponent $n$ is a parameter to the class, \texttt{store\_number} is of type $\RR \rightarrow \RR$ and writes to $x$; the method \texttt{compute\_power} is of type $\RR\rightarrow \RR$ and reads its input from state variable $x$. The ``obvious'' implementations differ by choices of the parameter $n$, for example
\begin{table}[H]
\begin{tabular}{l l c l}
	\multicolumn{4}{l}{\texttt{object Squarer}}\\
    \quad \texttt{params} & $n = 2$\\
	\quad \texttt{methods} & \texttt{store\_number} & : & $z\mapsto z$ \\
	\quad & \texttt{compute\_power} & : &  $z\mapsto z^n$  \\
\end{tabular}
\end{table}
is an \emph{inhabitant} of this class type which computes the square, and it is an \emph{instance} of the class
\begin{table}[H]
\begin{tabular}{l l c l}
	\multicolumn{4}{l}{\texttt{class PowerComputerClass}}\\
    \quad \texttt{params} & $n : \ZZ$\\
	\quad \texttt{state} & $x$  & : & $\RR$\\
	\quad \texttt{methods} & \texttt{store\_number} & : & $z\mapsto z$ \\
	\quad & \texttt{compute\_power} & : &  $z\mapsto z^n$  \\
\end{tabular}
\end{table}
where in said instance the parameter $n$ is set to the value $2$. We can thus write, formally, \texttt{Squarer} : \texttt{PowerComputer}; also, in common object/instance constructor syntax, one has \texttt{Squarer := PowerComputerClass}($n$ = 2). Note the difference between object, class, and class type: an object is an \emph{instance} of a class, both of which have the same class type, \emph{inhabited} by the object (= value). A class type does not specify a particular implementation as a class, multiple classes may inhabit the same type. There may be multiple distinct instances of a class, each with their own parameter values and states. 

We are now ready to sketch how formal typing simplifies specification in the two ML examples. In the ``pipelines'' example (e.g.\ of pipelines as found in scikit-learn), the key objects are ``transformers'' and ``supervised learners''. We work through the ``supervised learner'' example for illustration. First, we defined a suitable class types:
\begin{table}[H]
\begin{tabular}{l l c l}
	\multicolumn{4}{l}{\texttt{class type SupervisedLearner}}\\
    \quad \texttt{params} & \texttt{paramlist} & :& \texttt{paramobject}\\
	\quad \texttt{state} & \texttt{model}  & : & \texttt{mathobject}\\
	\quad \texttt{methods} & \texttt{fit} & : & $(\calX\times \calY)^N\rightarrow \texttt{model}$ \\
	\quad & \texttt{predict} & : &  $\calX \times \texttt{model} \rightarrow \calY$  \\
\end{tabular}
\end{table}
considered a parametric type in $N, \calX, \calY$, and where \texttt{paramobject} and \texttt{mathobject} are types of abstract representation of parameters and model objects respectively (e.g.\ umbrella types).

The class type is not yet the full scitype, because:
\begin{itemize}
\item Conceptually, supervised learners typically can be applied to multiple choices of $N, \calX, \calY$
\item Classifiers and regressors are defined by specific choices of $\calY$
\item Admissible input data types $\calX$ define specific sub-types of supervised learners
\end{itemize}

We can use the class type \texttt{SupervisedLearner} to define the scitypes:
\begin{itemize}
\item A formal object $x$ is of scitype ``supervised classifier on primitive features'' if it is of the type that is the type union over \texttt{SupervisedLearner} over the type $\calY$ being any finite set, the type $\calX$ a finite product of primitive scitypes (categoricals, numericals), and $N$ a natural number.
\item A formal object $x$ is of scitype ``supervised regressor on primitive features'' if it is of the type that is the type union over \texttt{SupervisedLearner} over the type $\calY$ being any measurable sub-set of $\RR$, the type $\calX$ a finite product of primitive scitypes (categoricals, numericals), and $N$ a natural number.
\end{itemize}

Inhabitants of the scitype ``supervised classifier on primitive features'' are specific algorithms following the specified interface -- for example, the following implementation of ``always predict the majority class on the training set'':
\begin{table}[H]
\begin{tabular}{l l c l}
	\multicolumn{4}{l}{\texttt{object MajorityDummyClassifier}}\\
	\quad \texttt{state} & \texttt{model}  & : & $\calC$\\
	\quad \texttt{methods} & \texttt{fit} & : & $\left((x_1,y_1),\dots,(x_N,y_N)\right)\mapsto \underset{c\in \calC}{\arg\!\max} \sum_{i=1}^N \mathbbm{1}[y_i = c]$ \\
	\quad & \texttt{predict} & : &  $(x,c)\mapsto c$  \\
\end{tabular}
\end{table}
where this object has no parameters, i.e.\ an empty parameter list. Note that this is not the only possible implementation of ``always predict the majority class on the training set'', since instead of only storing the majority class in \texttt{fit}, one could store all the training data and compute the majority class later in \texttt{predict}. Scitypes for transformers can be defined similarly; we defer to that until section \ref{sec:implementation}.

We conclude with pointing out some important conceptual distinctions related to the concept of scitypes:
\begin{itemize}
\item Between a class in the OOP sense, and a class type. A ``class'', as in Python, usually already contains blueprint implementations for some methods; whereas the class type does not specify an implementation for the method, and only records its input-output type and state access signature.
\item It is important to understand that type systems usually present in programming languages (e.g.\ float, string, etc.), sometimes called machine types, are not the same as scitypes, nor are they sufficient to inform ML design. As argued above, the reason is that types in programming languages do not capture statistical properties which are crucial to identify and abstract elements in the ML domain. Rather, scitypes are separate or orthogonal to the type systems typically found in programming languages.
\item We would also like to note that a scitype is not “meta-data”, i.e.\ data about data but rather a property of the data.
\end{itemize}

Having defined scitypes, we are now in a position to see how scitypes formalize the mapping of mathematical objects from the ML domain onto classes and operations in a programming language. In the next section, we will derive key design principles based on our conceptual analysis and scitypes in particular.

\subsection{Prior art}
We are unaware of literature instances of scitypes for learning strategies, or a mathematical formalization in the context of type theory. The closest on the formalism side is perhaps the area of type theory for structured objects \cite{Pierce2002}. On the conceptual side, scitypes tend to be a frequent, implicit concept appearing in discussion of algorithm types, especially in the toolbox communities. On the software side, scitypes for data set representations (e.g.\ scientific column types of data frames), have been an increasingly common idea, for example data frame column typing in base R \cite{RCoreTeam2014}, explicit scitype hints in dabl\footnote{\url{https:\\github.com/amueller/dabl}}, and perhaps most stringently and explicitly realized in the parallel type system implemented by the \software{ScientificTypes.jl} module of the \software{MLJ} ecosystem \cite{Blaom2020a}, which also introduces the shorthand term ``scitype''.
Implicit scityping of algorithms can be thought of as conceptually guiding behind many existing toolbox designs (e.g.\ \software{scikit-learn} \cite{Buitinck2013}, \software{mlr3} \cite{lang2019mlr3}, \software{MLJ} \cite{Blaom2020}). Though we are unaware of a fully explicit implementation of a full (sci)type system. While at the time being it therefore only has hypothetical status, one may conjecture that implementations may emerge as the ML software designs matures further.

\section{Design principles}
\label{sec:principles}
Having introduced our conceptual model, we now aim to derive practical design principles that can be used to develop ML toolboxes. In particular, we propose three guiding principles for ML toolbox design:
\begin{itemize}
    \item \textbf{Architectural separation of key conceptual layers} for data, algorithmic learning strategies, tasks, and workflows -- this is a special case of the common ``separation of concerns'', ``layering'' or ``decoupling'' design principles, applied to the conceptual categories previously outlined.
	\item \textbf{Scitype-driven object and interface model}, especially algorithmic learning strategies and mathematical objects (e.g.\ data sets, distributions, models). Interface points should mirror the conceptual abstractions of defining character for the scitype. For example, parameter setting, fitting and prediction defining a supervised prediction strategy and its interface. This can be seen as a special case of the ``encapsulation'', ``modularity'' and ``interface abstraction'' design principles.
	\item \textbf{Declarative syntax and symbolic specification}, motivated by the scitype formalism and minimizing boilerplate code. For example, specification of supervised learning pipelines should be as close as possible to their mathematical definition in the scitype formalism. This can be seen as a special case of ``symbolic abstraction'' and ``language abstraction'' design principles.
\end{itemize}

We proceed by laying out more details on each of these principles and how they are obtained from the conceptual model, using some of the running examples.

\subsection{Architectural separation of conceptual layers}
\label{sec:principles.layers}

To identify the right architectural layers means identifying key abstraction points -- i.e.\ parts of the use case that may be ``switched out'' while others are kept constant. For example, the same algorithm can be applied to different data sets, or two algorithms of a similar kind may be applied to the same data set. Thus, both algorithms and data sets should be abstraction points.

In this vein, based on the conceptual model one can identify four key abstraction points and thus architectural layers:
\begin{itemize}
\item Data sets and data related concerns; representation and manipulation of concrete data tables. For example, the concept of a data frame and operations on it.
\item Algorithmic learning strategies and related concerns; specification of concrete learning strategies, and learning strategy specific operations such as data ingestion or application of the algorithm. For example, the concept of a specific supervised prediction algorithm, such as a random forest, including functionality for fitting and prediction.
\item Learning tasks and related concerns; specification of the aim of learning, performance indicators. For example, the supervised learning task in abstraction, or in concrete reference to a data set and specific performance metrics such as the mean squared predictive error.
\item Workflows and related concerns; as special cases, workflows for evaluation, inspection, deployment or monitoring, with key concerns being specification and execution. For example, the concept of a predictive benchmarking experiment, or model interpretability diagnostic.
\end{itemize}

We postulate that these layers should be the basis for abstraction within and separation between, since they delineate ``independent moving parts'' in the conceptual model. While there are some intrinsic conditionalities and dependencies between the layers (e.g.\ only certain workflows make sense for supervised learning algorithms or tasks), varying parts from key use cases are typically localized within just one of the layers, e.g., choice of data set, choice of learning strategy, etc -- in line with our conceptualization of the task in section \ref{sec:cmodel.task}.

We derive some important consequences from this postulate, to address some common pitfalls or anti-patterns\footnote{Anti-patterns refer to common but poor solutions to a recurring software design problem in the spirit of \citet{brown1998anti}.}
\begin{itemize}
	\litem{Separation of learning strategies and data.} Models can be fitted to data or applied to data, but specification of learning strategies should be separate from how the data is specified. Algorithmic application of learning strategies, such as fitting or prediction, should be separate from operations intrinsic to the data.
	\begin{itemize}
        \item Important consequence 1: Software architecture of learning strategies should be abstract in the sense that it can be applied to ``data sets of a certain kind'', e.g.\ data frames.
        \item Anti-pattern 1: Writing code where the learning strategy is specific and hence inseparable from a particular data set. The level of generality depends on the use case, but it usually should be substantially beyond a ``data frame with the same column names''.
		\item Important consequence 2: Data cleaning operations sit in the data layer, data transformations for the purpose of modeling sit in the learning strategies layer. This may put the same operation in either depending on purpose. For example, record removal for cleaning purposes to create a consolidated data set should sit in the data layer. Record removal to prepare fitting of a prediction strategy should sit with that strategy. Missing data imputation, for the purpose of applying an algorithm that does otherwise not support missingness, should usually be part of a learning strategy pipeline that includes the imputation, as the purpose is not intrinsic to the data.
        \item Anti-pattern 2: Conflating concerns of modeling, tuning or model evaluation with data preparation, e.g.\ locating imputation or feature extraction for the purpose of modeling not with modeling but with data loading; or, making the creation of training/test folds for tuning an algorithm part of the data loading and cleaning workflow. It is worthwhile noting that the software engineering anti-pattern of conflating conceptual and architectural layers also easily enables common methodological mistakes along the lines of information leakage.
	\end{itemize}
	
	\litem{Separation of learning strategies and task specification.} Fitting and application of learning strategies requires task information (e.g.\ target variable, forecasting horizon), but strategy specification should be separate from task specification.
	\begin{itemize}
		\item Important consequence 3: Software architecture of learning strategies should be abstract in the sense that it can be applied to ``tasks of a certain kind'', e.g.\ general supervised prediction tasks, or forecasting tasks. For example, a general supervised learner should be able to take data sets with any feature/label columns; a forecaster should work with a number of different forecasting horizons. As passing concrete data sets, this information should be recognized by, but extrinsic to the learning strategy.
		\item Anti-pattern 3: An architecture that fails to separate or enforces specification of learning strategies and tasks at the same time. For example, requiring that the parameters of an SVM be specified together with the column names to be predicted, that is, an architecture where representation of an abstract learning strategy is not possible without a concrete specification of how it is applied to a specific learning problem.
	\end{itemize}
	
	\litem{Separation of learning strategies and workflows such as performance evaluation and monitoring.} A learning strategy may be deployed in accordance with a concrete task or within a workflow, but the workflow in which a learning strategy is used should be separate from the learning strategy itself.
	\begin{itemize}
		\item Important consequence 4: Concerns specific to advanced model performance evaluation workflows (e.g.\ train-test splits, cross-validation, loss functions) should not be considered part of learning strategies or their design. As before, the location of a specific methodological concept should be motivated by the purpose. For example, re-sampling and loss estimation for the purpose of model tuning is part of a learning strategy, thus should be part of the learning strategy layer. By contrast, re-sampling and loss estimation for the purpose of performance evaluation is with the purpose of evaluation, thus should be part of the workflow layer.
		\item Anti-pattern 4: Conflation of learning strategies and performance evaluation, e.g.\ a design where evaluation and tuning are considered identical. As in anti-pattern 2, the design mistake facilitates (but is not identical to) common methodological mistakes, such as evaluating on the training set, or conflating validation and test splits.
		\item Important consequence 5: Concerns related to monitoring and scrutiny of a learning strategy (e.g.\ logging, diagnostics) should be separate from and external to the learning strategy.
        \item Anti-pattern 5: The ``learning strategy that scrutinizes itself'', conflating the concern to carry out learning with the concern to check whether or how it is carried out. As in anti-pattern 2 and 4, the design mistake facilitates common methodological mistakes, such as circular reasoning in certifying reliability of a learning strategy.
	\end{itemize}
\end{itemize}

\subsection{Scitype-driven encapsulation}
\label{sec:principles.scitype}

The next set of design principles concerns object-level architecture. Our key postulate is that the design of architectural units, especially classes and objects, should be driven by scitype, as identified in conceptual analysis, and formally defined for algorithmic and mathematical objects. The main reason for postulating this design principle is its centrality in our conceptual model, which we argue is a formalization of typical data scientific conceptualization. According to domain-driven design \cite{Evans2004}, such conceptual lines should guide architectural decisions, especially if objects are delineated as mathematical or algorithmic units. Following this reasoning, the postulated correspondence between scitype and architectural units should be taken into account for:
\begin{itemize}
\item Definition of modules, classes and objects; especially when considering which functionality to ``bundle''
\item Interfaces of modules, classes, and objects; especially when considering key interface points
\item State change schemas of entity objects and classes implementing entity objects
\end{itemize}

We derive some important consequences for architecture in languages with class/object features:
\begin{itemize}
	\litem{Classes, class interfaces, and class templates as the simplest possible representation of scitype.} Given scitypes identified in the conceptual model, classes and class templates should be primarily considered as implementation of scitype inhabitants. A scitype defines a class interface that multiple classes can follow, ensuring non-proliferation of classes and templates.
	\begin{itemize}
		\item Important consequence 1: Key interface points in the conceptual model or user journey define the class architecture for instances of a given scitype. This can be combined with inheritance or template patterns (see section \ref{sec:patterns.ML} below). For example, the scitype of a supervised learning algorithm maps onto the familiar \texttt{fit}/\texttt{predict} interface in \software{scikit-learn}.
		\item Anti-pattern 1: Individual objects of the same scitype being implemented in separate, disconnected parts of the architecture; or, interfaces that are inconsistent between different algorithms of the same scitype.
        \item Important consequence 2: Scitypes defining entity objects with non-trivial state changes should map onto classes, interfaces, or other architectural motifs that ensure that the state and its change remain attached with the ``entity''. For example, a supervised prediction algorithm retains its identity when fitted because the identity as an entity is part of the conceptual model.
        \item Anti-pattern 2: In implementation architecture, the different states of entity objects are separated. For example, fitting a supervised prediction algorithm produces a ``fitted model'' which does not allow to trace back its entity lineage to the model and parameter specification that is fitted.
	\end{itemize}

	\litem{Operations with conceptual objects are methods of classes, operations on conceptual objects are operations on classes.} Operations \emph{of} conceptual objects (e.g.\ model fitting or application of a fitted model) should be methods of the class implementing that conceptual object. Operations that work \emph{on} conceptual objects (e.g.\ pipeline building) should be operations on (or with) classes.
	\begin{itemize}
		\item Important consequence 3: During conceptual analysis, use cases, user journeys, and mathematical definitions should be scrutinized for which operations of an object are key, in the sense of being a defining property of its scitype. In the architecture design, these operations should be methods of the class that implements the object, or otherwise closely coupled to the object (such as by being a queriable part of the same module). For example, fitting a model should be a method of the class that implements the model.
		\item Anti-pattern 3: Representation of conceptual objects is architecturally uncoupled from functionality that is considered ``part'' of it in the conceptual model. For example, an architecture where it is impossible to find the function that fits a supervised prediction model by application of code, given only the supervised prediction model.
	\end{itemize}

	\litem{Important operations on objects with a scitype define a higher-order scitype.} A higher-order scitype should be considered in design if it re-occurs in the use case. For example, composition of individual learning strategies to a pipeline, tuning or ensembling. A higher-order algorithm scitype found during conceptual analysis should be considered a first-class citizen, e.g.\ in terms interface design and implementation via classes.
	\begin{itemize}
		\item Important consequence 4: A good design will treat ``atomic'' (zeroth-order) scitypes (such as supervised learners) and ``higher-order'' scitypes (such as composition and tuning meta-strategies) consistently. In particular, class and interface designs will be consistent, inter-operable, and compatible between ``atomic'' scitypes and ``higher-order'' scitypes.
        \item Anti-pattern 4: Higher-order scitypes being second-class citizens, e.g.\ through implementing atomic scitypes as classes, while implementing higher-order scitypes (like tuning) as functions; or, lack of encapsulation and defined interfaces in higher-order scitypes.
	\end{itemize}

\end{itemize}
We would like to point out that on first sight, some of our stated consequences may seem to exclude functional programming paradigms and favor imperative (classical) object orientation, e.g.\ insisting on lineage of entity objects or methods. This is not true. While it is less enforced by other languages, entity lineage (the object remembers its ``identity'') and method lineage (methods ``belonging'' to an object can be queried) is implementable in functional object orientation paradigms such as in R or Julia, e.g.\ typed dispatch systems \cite{Chambers2014}.

\subsection{Declarative syntax and symbolic specification}
\label{sec:principles.symbolic}

With the scitype formalism readily available, it is a natural idea to leverage it for architectural design, as discussed above. We go one step further and postulate that scitype should determine syntax of usage, in a notebook-style user journey or in code development. Most importantly, this should cover the following instances:
\begin{itemize}
	\item Specification of formal algorithmic or mathematical objects, especially of parameter settings, composites, and higher-order constructs
    \item Execution or application of formal algorithmic or mathematical objects, especially when using scitype-defining methods
    \item Inspection of formal algorithmic or mathematical objects, especially when inspecting parameters, properties, or state
\end{itemize}

We motivate this design principle from domain-driven design, more specifically the principle that architecture should follow the conceptual model \cite{Evans2004}. Since formal algorithmic or mathematical objects are key concepts in the conceptual model, and manipulation of said objects arises in key interface points and use cases, it follows that the instructions for such manipulation should be as close to the formal conceptualization as possible -- in terms of mathematical objects and scitypes.

We list a few requirements arising from this high-level principle:
\begin{itemize}		
	\litem{Encapsulation of low-level functionality.} The highest level of code abstraction should be at least at the level of abstract mathematical and algorithmic objects, and scitypes. It should not be necessary to access or configure specifics of implementation. This is a special case of the ``no boilerplate code'' design principle, where, in accordance with the high-level principle, boilerplate is defined as being lower-level than the key conceptual objects.
	\litem{Parsimony of syntax.} Syntax of specification and usage should be as simple as possible, while being at a sufficient level of abstraction. This is a special case of the ``simplicity'' design principle.
	\litem{Congruence of syntax and conceptual semantics.} Syntax of specification and usage should follow, as closely as possible, the conceptual model of formal algorithmic and mathematical objects, especially the scitype formalism.
	\litem{Consistency of syntax across objects.} Syntax of specification and usage should be consistent across different formal algorithmic and mathematical objects, and across different levels of abstraction. This should in particular be the case for features or properties shared across objects or layers, such as the representation of parameters, states, or methods, across objects with different scitypes.
\end{itemize}

We will use these requirements in developing our generic design patterns in the next section. 

\subsection{Prior art}
As stated in the introductory section \ref{sec:intro}, we are unaware of a literature reference that formulates re-usable design principles specific to ML toolboxes, and systematically ties these to concrete design patterns. To our knowledge, there is also only a single references discussing high-level principles in the context of the design of a specific toolbox, \cite{Buitinck2013}, which contains some of the consequences discussed here; compare Section~\ref{sec:principles.scitype} consequence 1 with ``non-proliferation'', consequence 3 with ``composition'', and layer separation in Section~\ref{sec:principles.layers} with the banana/gorilla/jungle metaphor , and ``consistency'' in Section~\ref{sec:principles.symbolic} with ``consistency''.

\section{Design patterns}
\label{sec:patterns}

Having derived a set of design principles, we are now in a position to propose re-usable design patterns building onto the well-known patterns in \citet{Gamma2002}. Using object-oriented programming (OOP), each design pattern motivates, describes and names an important recurring design in ML software. Design patterns make it easier to reuse successful designs and architectures by making them more accessible to developers of new toolboxes.

We start by reviewing core concepts of OOP and argue that object orientation is a suitable language paradigm to implement patterns in accordance with the design principles outlined in the previous section. We then develop key design patterns for the general data scientific setting, drawing on object-oriented patterns as described for example in \citet{Gamma2002}.

In the next section, we illustrate these patterns by examples from contemporary practice and provide a collection of implementation blueprints.

\subsection{The case for object orientation in AI architecture}
\label{sec:patterns.oop}

A core architectural principle -- which has already stood the test of time as part of the existing major AI toolboxes\footnote{To name a few: \software{Weka} \cite{Hall2009}, \software{mlr3} \cite{lang2019mlr3}, \software{scikit-learn} \cite{Buitinck2013}.} -- is that of \emph{representing learners as objects}, in the sense of the programming paradigm commonly called ``object orientation'' or ``object oriented programming'' (OOP). Arguably, one may also consider this paradigm to extend more generally, to data or mathematical objects in the sense of section \ref{sec:cmodel.mathobj}.

OOP is the predominant paradigm for complex ML/AI software and often assumed without rationale. In this section, we provide a brief overview of the core characteristics of OOP, and more importantly, give key reasons why OOP is especially well suited for constructing ML/AI toolboxes.

OOP is a general, not ML-specific motif supported by a programming language, and may take various forms depending on the host language. For example, it may be of imperative, functional or mixed flavor. Some examples relevant for data science are: pure class-based OOP in Java or Python; functional, dispatch-based OOP in Julia or Go; and the various mixed OOP paradigms in R (S3, S4, R6).

Defining features of OOP are:
\begin{itemize}
\item \emph{Structured objects} with \emph{variables}, e.g.\ a ``duck'' type object which may contain object (state) variables such as \texttt{number\_of\_feathers}, or \texttt{is\_dead};
\item \emph{Methods}, which are functions specific to, and/or belonging to the aforementioned structured objects; e.g.\ a function \texttt{quack}, \texttt{waddle}, or \texttt{remove\_one\_feather}, which may mutate internal states, and/or which may return an output.
\item \emph{Distinction of classes and objects}, with a ``class'' being a generic blueprint for a concrete ``object'', where the class-object relationship is similar to the type-variable relationship, e.g.\ type being \texttt{integer} vs a concrete \texttt{integer} typed variable with value $42$. For example, there may be multiple concrete ``object'' instances \texttt{Huey}, \texttt{Dewie}, \texttt{Louie} of a generic class called \texttt{duck}.
\item \emph{Class inheritance} (or type ordered dispatch in functional OOP) and \emph{object type polymorphism}. Informally, inheritance allows us to model a containedness hierarchy of classes, e.g.\ \texttt{duck} being a ``sub-class'' (i.e.\ a kind) of \texttt{bird}. Object type polymorphism means that methods, object variables, and functions for classes defined as a sub-class will automatically default to those of the super-class, unless overwritten. e.g.\, it makes sense to define generic \texttt{number\_of\_feathers} and \texttt{waddle} for a class \texttt{bird}, then sub-class \texttt{duck} and \texttt{penguin}, where \texttt{quack} is implemented only for \texttt{duck}.
\end{itemize}

Optional, common features of OOP are:
\begin{itemize}
\item Class-specific \emph{constructor} and \emph{destructor} methods, called for creation and upon destruction of an object;
\item \emph{Strong typing}, \emph{weak typing} or \emph{duck typing}, depending on whether and how a class is considered as a type according to the language's type system;
\item \emph{Distinction of public, private or protected} for variables and methods, depending on whether access is possible from outside an object or not;
\item \emph{Distinction of static vs mutable} objects, variables or methods, depending on whether they may be changed after construction of the object;
\item \emph{Distinction of class variables/methods vs object variables/methods}, depending on whether they are common to all objects of a class, or only specific to an object instance;
\item \emph{Multiple inheritance}, \emph{decorators} and \emph{mix-ins}, depending on whether a class may simultaneously inherit from multiple super-classes, or have behavior modified by subsidiary templates;
\item \emph{Multiple dispatch}, depending on whether methods may dispatch on more than one inheritance/order hierarchy.
\end{itemize}

Most of the optional features may be mimicked by architectural patterns in an OOP paradigm with the defining features, even if not available as part of the host language. Examples are adoption of a naming convention for variables or methods, to signify private/public access; or, to signify constructor/destructor behavior for methods.

We argue that OOP is the natural software engineering paradigm for an architectural framework for ML/AI. The typical use case for OOP is for a conceptual model involving objects which
\begin{itemize}
\item may exist in \emph{multiple instances} (objects) following a common blueprint (class), e.g.\ instances 1 and 2 of linear model on data batch A; instance 3 of linear model on data batch B; or different instances of an entity object, such as different instances of probability distributions
\item possess \emph{multiple internal states} between which they can transition, e.g.\ newly set-up model vs fitted model after data ingestion
\item are \emph{subject to a defined collection of actions or methods} run repeatedly within, or across, instances -- e.g.\ ingesting data for fitting or predicting with a supervised learner
\item \emph{interact via defined interfaces} with each other, e.g.\ supervised learner object ingesting data from a data container object using a \texttt{fit} method; or AI pipeline object being composed of learner objects
\end{itemize}

Table \ref{tab:caseforoop} 
collates further parallels between natural features of the introduced conceptual models and features of OOP.

\begin{table}[ht]
	\centering
	\small
	\caption{\label{tab:caseforoop} Schematic mapping of typical OOP features onto AI/ML objects}
	\begin{tabularx}{\textwidth}{lXX}
		\toprule
		OOP feature & Ducks and birds & AI and learning machines \\
		\midrule			
		Class vs object &  blueprint class \texttt{duck} vs concrete objects \texttt{Huey}, \texttt{Dewey} and \texttt{Louie} & e.g.\ blueprint class \texttt{SVM} vs objects \texttt{my\_SVM1},  \texttt{my\_SVM2} that can be fitted to different data\\
		Methods & \texttt{walk}, \texttt{quack} & \texttt{fit}, \texttt{predict}, \texttt{inspect} \\
		Inheritance & \texttt{duck} sub-classes \texttt{bird}, which implements \texttt{remove\_one\_feather} & \texttt{SVM} sub-classes \texttt{supervised\_learner}, which implements \texttt{set\_params}\\
		constructor & setting up a \texttt{duck} & setup learner structure and parameters \\
		Class variables & properties of every \texttt{duck} e.g.\ swims, can quack & properties of every (sup.) \texttt{SVM} e.g.\ supervised, kernel method \\
		Object variables & properties of \texttt{duck} instances e.g.\ \texttt{number\_of\_feathers} & Fitted model state and hyper-parameters, e.g.\ weights, regularization parameters\\
		Mutable classes & ducks (can ingest bread) & learners (can ingest data)\\	
		Static classes & (all ducks are mutable) & formal mathematical objects e.g.\ distribution, sampler \\
		Multiple inheritance & the Indian Runner duck implements \texttt{run} from \texttt{running\_animal} & \texttt{SVM} may sub-class supervised learners and also kernel learners for methods \\
		Object interaction & Indian Runner duck herd  & composite ML pipeline \\
		\bottomrule
	\end{tabularx}
\begin{minipage}{\textwidth}
	\emph{Notes}: The table describes typical OOP features (first column) on aspects of the conceptual models for AI objects introduced in sections \ref{sec:cmodel.task} and \ref{sec:cmodel.mathobj} (third column), aligned with a supporting correspondence based on the ``ducks'' metaphor (second column).
\end{minipage}
\end{table}

The main architectural patterns for AI, to be discussed in this paper, arise in the context of OOP models of the above flavor, as application of so-called architectural design patterns in OOP. These, together with some necessary formalism, will be further expanded upon in detail, in section \ref{sec:patterns.ML} and thereafter.

Generally, an ``OOP design pattern'' is the term given to any reusable solution for a common software design problem, using OOP. Table \ref{tab:patterndef} gives a brief description of generic OOP design patterns referenced in this paper. Readers are referred to \cite{Gamma2002} for a full glossary of these terms and examples.

\subsection{Design patterns for ML toolboxes}
\label{sec:patterns.ML}

We now present a number of general software engineering design patterns for ML toolboxes. Generally an ``OOP design pattern'' is the term given to any reusable solution for a common software design problem, using OOP. The patterns we propose in this section are derived from the high-level principles in \ref{sec:principles} and the conceptual model in \ref{sec:cmodel}, building onto the well-known patterns in \citet{Gamma2002}. In this sense, the proposed patterns are ``canonical'' as they are largely implied by the choice of formalization as a scientific type.

For our exposition on key patterns, we will recapitulate some basics of design patterns. Table \ref{tab:patterndef} gives a brief description of generic OOP design patterns referenced in this paper. For a more detailed but non-ML-specific exposition of theses patterns, readers are referred to \citet{Gamma2002}.
\begin{table}
	\centering
	\small
	\caption{Overview of key design patterns from \citet{Gamma2002}}
	\label{tab:patterndef}
	\begin{tabularx}{\textwidth}{llX}
		\toprule
		Pattern & Type & Definition \\
		\midrule
		Template & Behavioral & Defines a common interface pattern through an abstract template class and inheritance. \\
		Strategy & Behavioral & Defines a family of interchangeable algorithms, with the specific implementation to be selected at run-time. \\
		Visitor & Behavioral & Implements operations as visitor objects to be applied to other objects. \\
		Composite & Structural & Blueprint for composing a class built from individual ``component'' classes. \\
		Decorator & Structural & Adds responsibilities to an object dynamically; a type of wrapper. \\
		Adapter & Structural & Converts the interface of a class into another; a type of wrapper. \\
		Facade & Structural & Provides a high-level interface abstraction to modules or lower level implementation. \\
		\bottomrule
	\end{tabularx}
\end{table}

The patterns we present here are:
\begin{itemize}		
	\litem{Universal interface points for mathematical and algorithmic objects.} These are the interface points that all formal objects should share, based on our conceptual model.
	\litem{The scitype template/strategy pattern.} A templating pattern that can be used to define and support scitype-specific interfaces and implementations of concrete algorithms with that scitype.
	\litem{Composition pattern 1: The modification pattern.} Implements an operation that modifies an object while preserving the scitype, for example tuning.
	\litem{Composition pattern 2: The homogeneous composition pattern.} Implements an operation where multiple objects are combined into one, with a known resultant scitype, for example ensembling with fixed scitype.
	\litem{Composition pattern 3: Scitype-inhomogeneous composition patterns.} Implements an operation where multiple objects, possibly of different scitypes, are combined into one, possibly of different resultant scitypes depending on parameters or component scitypes, for example pipelining (chaining transformation and prediction algorithms).
	\litem{The contraction factory.} Implements bulk-conversion of a composite into an atomic (non-composite) object.
	\litem{The reduction/composition factory.} Implements bulk-application or transformation of a formal object to a closely related scitype.
	\litem{The co-strategy pattern.} An abstraction to be used for encapsulating instructions or operations in relation to formal objects.
\end{itemize}

We illustrate these data scientific design patterns with a selection of contemporary examples from \software{scikit-learn}\footnote{Throughout the paper, we rely on the latest release available at the time of writing: v0.24.}. Further implementation examples from existing toolboxes is given in section \ref{sec:implementation}.

\subsubsection{Mapping formal objects onto classes and objects}

Before discussing patterns, we introduce some terminology in this section that will allow us to discuss patterns more clearly. For clarity of terminology, we will use the term ``formal objects'' to refer to the conceptual level, while the use of ``class'' and ``object'' will refer to the software and implementation side.

We start by discussing how ``formal objects'', i.e.\ mathematical and algorithmic objects (e.g.\ distributions, supervised learners) from the conceptual model map onto classes and objects in the OOP paradigm. In line with the discussion in section \ref{sec:patterns.oop}:
\begin{itemize}	
    \litem{Objects as representations of concrete formal objects and their current state.} Concrete formal objects, e.g.\ the Gaussian distribution with mean 0 and variance 1, are represented by objects. If the formal object is an entity object, the object representation also carries the information of the current state, e.g.\ in the case of a fitted linear regression model, coefficients would also be represented in the (software) object.
    \litem{Blueprints for formal objects are represented as classes.} Objects representing concrete formal objects are instances of classes that represent blueprints of a ``kind'' of formal object and which are possibly parametric, e.g.\ Gaussian distributions with mean and variance being parameters. The class in question will contain full implementation of functionality for possible instances, including specifics of representation, creation, and state change.
\end{itemize}

For terminological clarity, we will refer to (software) objects as in the first bullet point as ``formal instances'' or ``formal instance objects''. We will refer to classes as in the second bullet point as ``formal concretes'' or ``formal concrete classes''.

This terminology will be useful for discussing architecture since not all concrete classes in the usual sense (instantiable classes with a full implementation) are concrete classes representing (conceptual) formal objects; similarly, not all classes are concrete, and not all classes representing (conceptual) formal objects are formal concretes. Further, not all classes occurring in a software architecture will be related to formal objects.

\subsubsection{Universal interface points}

The conceptual model as discussed in sections \ref{sec:cmodel.mathobj-sci} and \ref{sec:cmodel.ML-sci} implies a number of key interface points that implementations of any formal objects should have -- we call these ``universal interface points'' (universal for implementations of formal objects). We list the universal interface points in table \ref{tab:universal-interface-points}.

\begin{table}
\caption{Universal interface points for all formal objects}
\label{tab:universal-interface-points}
\centering
\small
\begin{tabularx}{\textwidth}{lX}
	\toprule
	Interface point & Description \\
	\midrule
	Specification & construction of the object, and specification of hyper-parameters, e.g.\ constructing a normal distribution with \texttt{mean}=0, \texttt{variance}=1 \\
    Scitype inspection & query that returns the scitype of the object, e.g.\ a distribution \\
    Domain inspection & query that returns the domain of the object, e.g.\ absolutely continuous distributions over the reals\\
    Parameter inspection & query that returns names, values, and type/domain of hyper-parameters, e.g.\ retrieving the value (=1) and type (real number) of \texttt{mean} \\
    Parameter setting & setting values of hyper-parameters, e.g.\ setting \texttt{mean} to 2 \\
    Trait/property inspection & query that returns names, values, and type/domain of traits and properties, e.g.\ kurtosis of the distribution\\
	Object persistence & saving and retrieving objects, e.g.\ storing the distribution once constructed \\
	\bottomrule
\end{tabularx}
\end{table}

We note that parameters and traits/properties are separate concepts by definition of the interface -- we explicitly require these concepts to be separated by the interface, due to them being separate in the conceptual model (rather than implementation necessity). Further, it is important that the respective universal interface point not only returns values, but also the set of parameters or traits -- this is an important requirement for later patterns (e.g.\ composition or reduction).

Entity objects have additional universal interface points related to handling of object state, which we list in table \ref{tab:universal-interface-points-entity-objs}. We intentionally do not include state change as a universal interface point for entity objects, since the number and quality of interface points will depend on the scitype of the object.

\begin{table}
\caption{Universal interface points for entity objects}
\label{tab:universal-interface-points-entity-objs}
\centering
\small
\begin{tabularx}{\textwidth}{lX}
	\toprule
	Interface point & Description \\
	\midrule
	State inspection & query that returns the state of the object, e.g.\ fitted or unfitted for a linear regression model\\
	State variable inspection & query that returns names, values, and type/domain of state defining variables, e.g.\ regression coefficients \\
	\bottomrule
\end{tabularx}
\end{table}

The implementation of this pattern is as follows:
\begin{itemize}		
	\litem{Use of the template pattern and inheritance for universal interface points.} Abstract functionality for universal interface points should be implemented by a ``universal base class'' template. Any class implementing a formal object should inherit from this (or a suitable) universal base class, in the process being endowed with functionality of the universal interface points. This follows the ``template pattern'' in \citet{Gamma2002}. For example, the universal base class would implement general parameter functionality or trait inspection. Depending on case or language, universal interface points for entity objects may also be part of that base class (and not used for value objects), or added by mix-in or decorator patterns.
	\litem{One method (at most) per universal interface point.} Universal interface points should map onto at most one core interface method -- e.g.\ there should be one method by which all parameters can be inspected, as opposed to one method per parameter. Specific scitypes may implement ``shorthand'' methods for individual parameters or traits, but the pattern requires a universal method per interface point.
	\litem{Universal method signatures.} Methods and usage of universal interface points is designed to be universal across formal objects of any scitype.
	\litem{Compositionality and templating.} Interface implementations need to satisfy the templating and compositionality requirements in the next two patterns.
\end{itemize}

\textbf{Example:} The universal base class of the \texttt{scikit-learn} toolbox is the \texttt{BaseEstimator} class. It implements some of our suggested universal interface points; for example, \texttt{get\_params} for parameter inspection, and generic functionality for construction and specification in the constructor (\texttt{\_\_init\_\_}). Notably absent from \texttt{BaseEstimator} are scitype inspection, or state variable inspection (inspection is object specific and added on descendant level, and not universal).

\subsubsection{The scitype template/strategy pattern}
\label{sec:patterns.ML.strategy}

The central pattern used in contemporary toolbox architectures, and one that is often thought of as ``defining'' for ``scikit-learn-like'' toolbox is the scitype template/strategy pattern. It can also be considered as the primary pattern following from the design principles in section \ref{sec:principles.scitype}. As already evident from the naming, this pattern follows two generic \pattern{template} and \pattern{strategy} patterns from \citet{Gamma2002}, in the specific context of scitypes.

The conceptual purpose of the scitype/strategy pattern is to provide both abstraction and conformity for formal instance objects of a given scitype, e.g.\ distributions, or supervised classifiers.

The pattern applies to formal concrete classes that implement specific formal objects of a particular scitype $S$, e.g.\ concrete probability distributions (e.g.\ Gaussian distributions) that follow the ``distribution'' scitype, or concrete learning algorithms (e.g.\ random forest classifier) that follow the ``supervised classifier'' scitype.

The basic implementation pattern is as follows:
\begin{itemize}		
	\litem{Use of the template pattern and inheritance for scitype specific methods.} Abstract functionality for universal interface points should be implemented by a ``scitype base class'' template. Any formal concrete of a scitype $S$ should inherit from the scitype base class for the scitype $S$, in the process being endowed with functionality specific to the scitype $S$, and interface patterns characteristic of the scitype $S$. This follows the ``template pattern'' from \cite{Gamma2002}. Alternatively, the template can be adhered to by a combination of implicit specification and inheritance applied to formal concretes.
	\litem{Use of the strategy pattern for formal concretes of the same scitype.} Formal concretes of the same scitype $S$ should behave exchangeably within key use cases and workflows, and implement the same interface that is necessary for key behaviors and interactions. For example, all supervised classifiers implementing \texttt{fit} and \texttt{predict} methods with the same signature which are called within the fitting/deployment use case. This is ensured by adhering to the ``strategy pattern'' from \cite{Gamma2002} in relation to the scitype base class.
\end{itemize}

In concrete implementation, the formal concrete classes should (compatibly) inherit both from the universal base class (see above) as well as adhere to a scitype specific strategy pattern. This can be handled in multiple ways where suitability may also depend on language idioms. For example, multiple inheritance, the scitype base class inheriting from the universal base class, or combination of inheritance and implicit convention.

\textbf{Example:} supervised classifiers and regressors (for i.i.d./tabular data) define two common scitypes; these are scitypes central to \texttt{scikit-learn}'s use cases. In \texttt{scikit-learn}, supervised classifiers and regressors adhere to the template/strategy pattern. Classifiers and regressors are required to implement scitype-defining \texttt{fit} and \texttt{predict} methods with a unified interface; functionality is enforced by post-hoc validity checks rather than explicit inheritance. There are also scitype specific base classes for regressors and classifiers, namely \texttt{RegressorMixin} and \texttt{ClassifierMixin}, but their remit is mainly accessory functionality (scoring) and scitype inspection (signposting by inheritance), rather than handling the \texttt{fit}/\texttt{predict} interface.

\subsubsection{Scitype composition patterns}

In this paragraph, we cover a number of construction patterns for use cases that involve implementations of operations on formal objects. The key concepts on the formal side that are covered by the patterns are:

\begin{itemize}
\litem{Modification:} Taking a formal object and building another formal object out of it, of the same scitype. Common examples are tuning or output thresholding.
\litem{Composition:} Taking multiple formal objects and using them to build another formal object. Common examples are ensembling or pipelining.
\litem{Reduction:} Taking one or multiple formal objects and building another formal object, of different scitype.
\end{itemize}

In the above, we call the formal objects out of which the result is built the \emph{components}, and the result a \emph{composite}.
If the composite's scitype varies dependent on the component scitypes, we call the operation \emph{output-polymorphic}, otherwise \emph{output-monomorphic}.
An operation can be, at the same, time a composition and a reduction, if the composite is of different scitype than at least one of the components (one may also define ``reduction in the strict sense'' as a composition where \emph{all} component scitypes are different from the composite scitype).

With formal scitypes as in section \ref{sec:cmodel.ML-sci}, we can give a more formal definition in terms of (sci-)type notation. The reader content with intuition as given above may skip the formalism.

{\bf Definition:} A monomorphic composition operation is an function $f$ of type $A_1\times \dots \times A_N \rightarrow B$, where $A_i$ and $B$ are scitypes. $A_i$ are called component scitypes, and $B$ is called the resultant scitype (of the composite) of the operation $f$ and of its type. The operation $f$ is called: a \emph{modification} if $N=1$ and $A_1 = B$; a \emph{reduction} if there is an index $i$ such that $B\neq A_i$.
A (general) composition operation is a scitype union of monomorphic composition operations, i.e.\ a function $g$ of the type $C_1\vee \dots\vee C_M$ where $C_i$ are types of monomorphic composition operations. $g$ is called output-monomorphic if the resultant scitypes of $C_i$ are all identical (varying $i$); it is called output-polymorphic if the resultant scitypes of the $C_i$ (varying $i$) are non-identical. {\bf Definition end.}

We present a basic implementation pattern for composites and reducts, which we will later vary depending on the type of composition (e.g.\ modification, reduction, output-polymorphism). In vanilla form, it applies to output-monomorphic composition operations only, and is built as follows:

\begin{itemize}		
	\litem{Implementation by a class of the resultant scitype pattern.} The composition operation is modeled by a concrete class which has the scitype of the resulting composite. e.g.\, thresholding a supervised regressor's output is implemented as a class that also follows the supervised regressor scitype and its scitype specific template/strategy pattern. This is in line with principles 3 and 4 in section \ref{sec:principles.scitype}.
	\litem{Composition and parameter setting at construction.} The components are passed to the composite as objects at construction or initialization, together with parameter settings specific to the composite, using a combination of the ``composite pattern'' and the ``factory pattern'' from \citet{Gamma2002}. This also leads to a first-order-like specification syntax in accordance with requirements in section \ref{sec:principles.symbolic}.
	\litem{Components are variables of the composite.} The components are treated as variables of the composite.
	\litem{Component state as part of the composite's state.} The components become part of the composite's state, potentially changing state whenever the composite changes state.
    \litem{Compositionality of universal and scitype specific interfaces.} Component interface points should be visible via the composite interface points, e.g.\ hyper-parameter interfaces, or state interfaces. This also puts additional requirements on the implementation of the universal interface points as well as scitype specific interface points, such as conventions on how to access hyper-parameters of components via the hyper-parameter interface of the composite, or how to list state variables of components via the state variable interface of the composite. This is to ensure concordance with principles 3 and 4 in section \ref{sec:principles.scitype}.
\end{itemize}

An interesting, and often misunderstood, consequence of the pattern is the ``currying out of method inputs'' -- that is, that inputs to component methods do not become part of the (software implementation of the) composite. For example, the operation of tuning a supervised learner, if very naively implemented, takes the supervised learner and some training/validation data on which the best parameters are determined; the output is the supervised learner whose parameter are set to the ``best parameters''. However, this is not following the composite pattern as outlined above: if it is correctly followed, the data is \emph{not} an input to the composition operation. Instead, tuning is performed inside the fitting method of the composite, which has access to the data in deployment. Since the data is passed to the fitting later anyway (namely: at deployment), there is no need to pass the data to the composition operation too.

We proceed with outlining some variation patterns depending on the nature of the composition operation.
\begin{itemize}
	\litem{Adaptor pattern for scitype change.} For polymorphic composition operations that are output-monomorphic, the adapter pattern from \citet{Gamma2002} is suitable, in the form of adapting potentially different input scitypes to the interface of the resultant scitype.
	\litem{Strategy dispatch for output-polymorphic composition.} If there can be multiple composite scitypes dependent on component scitype, this behavior can be modeled by strategy dispatch on input scitype. This can be achieved with a suitable application of the strategy pattern (selecting by component scitype), by run time decoration (dependent on component scitype), by method dispatch (on component scitype), or a suitable combination thereof.
	\litem{Wrapper and visitor patterns for simple modifications.} In the case of modification or reducts with a single component (or few fixed scitype component), wrapper and visitor patterns from \citet{Gamma2002} may be alternatives worthwhile considering.
\end{itemize}

\textbf{Example:} In \texttt{scikit-learn}, the tuning operation \texttt{GridSearchCV} and pipelining operation \texttt{Pipeline} adhere to the composite pattern. Both operations are output-polymorphic. \texttt{GridSearchCV} addresses the output-polymorphism by method polymorphism (the \texttt{score} method); \texttt{Pipeline} addresses the output-polymorphism by method dispatch (to \texttt{predict} vs \texttt{transform} dependent on resultant scitype).

\subsubsection{Compile-time composition and contraction}

There are situations in which operations may be preferable on the class level (i.e.\ statically) rather than on the level of objects or instances (i.e.\ at run-time), as in the composition patterns above.

The most common such situations are as follows:

\begin{itemize}
\litem{Modification or composition at compile-time.} Creating a class whose instances are composites as if obtained by the scitype composition/reduction pattern.
\litem{Contraction to an atom.} Turning a composite with specific parameter settings into a pre-defined object -- e.g.\ creating a class whose instances are fully specified supervised prediction pipelines with a given parameter specification.
\litem{Contraction of components.} Removing the component hierarchy while leaving parameters settable at construction at compile-time
\litem{Contraction of parameters.} Setting some parameters or components to specific values at compile-time.
\end{itemize}

The primary use cases for compile-time composition and contraction and not run-time composition are:

\begin{itemize}
\item leveraging blueprints to create other blueprints, e.g.\ bulk-converting loss functions for supervised regression to loss functions for supervised prediction of numerical-valued sequences or segmentation.
\item ``fixing'' a reference implementation for an important composite; e.g.\ for convenience in frequent re-use, transparency, reproducibility or replication; a common example is the common composite of a grid-search tuned support vector machines with prior input normalization, as the method does not tend to perform well in isolation.
\end{itemize}

We proceed with outlining some variation patterns depending on the nature of the composition operation.
\begin{itemize}
	\litem{Compile-time decoration.} This is using the ``decorator pattern'' or ``wrapper pattern'' from \citet{Gamma2002} (at compile-time), and is mostly suitable for operations with a single component, e.g.\ modification or simple reductions. The pattern is used to decorate or wrap the component on class-level.
	\litem{Compile-time factories.} This is using the ``factory pattern'' from \citet{Gamma2002} (at compile-time). A typical implementation would suitably construct the final composite at construction of the factory class and return itself.
\end{itemize}

\textbf{Example:} Any ``naive'' definition of shorthands for a frequently used composites is compile-time composition. Applying standard reductions such as thresholding in probabilistic classifiers can be seen as an instance of compile-time reduction. \texttt{scikit-learn} does not seem to make widespread use of this pattern.

\subsubsection{The co-strategy pattern}

In handling implementations of formal objects, it may be useful to abstract recurring motifs in passing information, specifically in:
\begin{itemize}
	\litem{Object construction and parameter manipulation.} For example, setting the values of hyper-parameters at construction of a formal object, for example specifying the kernel function in a support vector machine.
	\litem{Passing arguments to scitype specific interface methods.} For example, specifying task information such as variables to predict, or a forecasting horizon, to the \texttt{fit}/\texttt{predict} methods of a learning algorithm.
\end{itemize}

The pattern relies on conceptualizing the recurring motif as an object in its own right, with its own scitype -- notably, not from first principles, but in secondary relation to the ``primary object(s)'' from which it is abstracted.
Implementation, in consequence, relies on:

\begin{itemize}		
	\litem{Use of the strategy pattern for the motif,} considered in its own right, and the scitype template/strategy pattern to implement a ``conceptual kind'' of motif. For example, specification of an abstract learning task in relation to a data set on which a learning strategy is employed, or specification of cross-validation parameters used in tuning.
	\litem{Methods implement operations intrinsic to the motif.} Functionality that is specific to the motif without the context should be located with the motif implementing class rather than the primary object. For example, if the motif abstracted is a data transformation, functionality to execute the data transformation should be localized with the specification of the data transformation.
	\litem{Consistency of co-strategies.} If the same conceptual motif occurs as parameters or arguments in the context of multiple operations, it should be handled by the same co-strategy or co-strategy scitype. For example, if cross-validation specifications appear in tuning operations and evaluation workflows, they should be handled by the same class or scitype, as opposed to one class for tuning and one class for evaluation.
\end{itemize}

On a side note, co-strategies often implement what one may conceptually a ``specification'', for behavior or an algorithm -- however, we expressly avoid that term since ``specification pattern'' is an established design pattern with incongruent meaning.

\textbf{Example:} Cross-validation specifications (such as the cross-validators in the \texttt{model\_selection} module of \software{scikit-learn}) can be seen as a co-strategy obtained from encapsulating the ``specify the parameters for cross-validation'' in tuning via \texttt{GridSearchCV}. Transformers, a widely used conceptualization of data manipulation operations (such as feature extractions) used as part of a pipeline composite, can be considered a co-strategy obtained from encapsulating prior data manipulation steps in the modification that is ``first do the data manipulation, then apply the supervised learner''.

\subsection{Prior art}

To our knowledge, this content is entirely novel in the sense of abstract and systematic formulation, while drawing inspiration from the non-ML-specific work of \citet{Gamma2002}, and domain-driven design practice put forward in \citet{Evans2004}. Many of the patterns presented here have been implicitly used in the design or implementation of existing state-of-art toolboxes (also see examples above), especially \texttt{Weka} \cite{Hall2009}, \texttt{scikit-learn} \cite{Pedregosa2001, Varoquaux2015, Buitinck2013}, \texttt{mlr3} \cite{lang2019mlr3, Bischl2016}, and \texttt{MLJ} \cite{Blaom2020a} -- but not named or explicitly identified as generic patterns. The co-strategy pattern, and especially the encapsulation of task information into objects, is partially inspired by the APIs of \software{mlr} \cite{Bischl2016, lang2019mlr3} and the \software{openML} platform \cite{Vanschoren2014}.

\section{Selected implementation examples}
\label{sec:implementation}

In this section, we illustrate how our proposed principles and patterns from section \ref{sec:principles} and \ref{sec:patterns} apply to concrete interface designs. We focus on learning algorithms as the key objects in typical ML workflows, including predictors, transformers and composite models such as pipelines. We illustrate not only how our analysis can explain central aspects of contemporary toolboxes. We will use \software{scikit-learn} as our running example, but similar designs can be found in other toolboxes such as \software{mlr3} \cite{lang2019mlr3}, \software{MLJ} \cite{Blaom2020a} or \software{Weka} \cite{Hall2009}.

Many of the proposed designs are not new, but they are for the first time derived from general design patterns. Through our systematic approach, we also identify a number of novel aspects which do not seem to be covered by contemporary state-of-art designs, such as unified handling of fitted parameters for AI algorithms. In addition, we will illustrate how our analysis can guide the design of new toolboxes using some examples from our work on \software{sktime} \cite{Loning2019}, a new unified toolbox for ML with time series.

In general, we proceed in our designs as follows:
\begin{itemize}
	\item As argued in section \ref{sec:cmodel.task}, any interface design for algorithms starts with a problem specification. The algorithm scitype is then defined as the counterpart to the task. The formal problem specification, or task, defines what to solve -- the precise ML problem in question -- in contrast to the algorithm scitype which defines how to solve it. Mathematically, the task can be understood as the codomain of the algorithm scitype. While algorithms are functions that, given some data, fit and return a function, the task defines the set of destination into which all of the returned functions of the algorithm are constrained to fall.
	\item In practice, a task specification also contains the information necessary to specify a particular learning task in regard to some practical problem at hand. For example, in supervised learning, the task contains at least the name of the target variable. In forecasting, it contains at least the forecasting horizon, that is the time points one wants to predict.
	\item From the task, we can then derive an algorithm scitype. As discussed in section \ref{sec:cmodel.sci}, a scitype defines both an abstract class interface and key statistical properties. Following the scitype template/strategy pattern defined in section \ref{sec:patterns.ML}, each algorithm scitype is implemented as an abstract base class, while specific algorithms are implemented as concrete classes inheriting the interface specification from the abstract base class.
	\item The template/strategy principle ensures that many ML models which differ only with respect to their internal algorithm but have the same purpose, also share the same interface. It separate the ``what'' from the ``how'', and encapsulates the intricacies of the solution (``how'') behind a clear interface (``what''), in line with the idea of ``intention-revealing'' interfaces from domain-driven design \cite{Evans2004}. This avoids exposing complex, model-specific implementations to users, and makes it easier to understand, compare and extend models.
	\item Scitype-driven encapsulation of algorithms also ensures that algorithms are decoupled from other common workflow elements such as data objects and other generated artifacts such as predictions, as discussed in section \ref{sec:principles.layers}.
	\item We give a list of key interface points for all learning algorithms, regardless of their specific scitype in table \ref{tab:learning-algorithm-interface}. Additional interface points for all scitypes are listed in table \ref{tab:universal-interface-points}.
	\item As argued in section \ref{sec:cmodel.objects}, algorithms are conceived of as entity objects. Entity objects characterized by having state changes, but with a thread of continuity that need to be maintained (e.g.\ fitted and unfitted model). Often the state spans several variables, and changes to those variables must be coordinated in order to maintain the consistency of the state. To manage the state changes through an interface, we conceive of learning algorithms as entity objects. As such, they possess multiple internal states between which they can transition, and states can be accessed and modified through several operations (e.g.\ fitting). In addition, the learning algorithm that is used during fitting is often tightly coupled to the prediction functional used during prediction. In other words, different prediction functionals require different learning algorithms. Construing learning algorithms as entity objects helps to maintain a close connection between the fitted prediction functional and the learning algorithm used to fit it. The additional interface points that we consider important for learning algorithms as entity objects are listed in table \ref{tab:universal-interface-points-entity-objs}.
	\item Traits of concrete algorithms can be captured by algorithm tags, implemented as class variables of concrete algorithm classes with shared functionality residing the abstract base class. Traits can be used for finer categorization of different algorithms of the same scitype. This is especially useful for testing where one may want to run the same tests on all algorithms of the same category. Traits are also useful for guiding users in algorithm choice, as they allow to query and filter algorithms by their traits. For example, one may define a trait for whether a supervised classifier can handle multi-label problems.
\end{itemize}

\begin{table}[tbh]
	\caption{Common interface points for learning algorithms}
	\label{tab:learning-algorithm-interface}
	\centering
	\small
	\begin{tabularx}{\textwidth}{lX}
		\toprule
		Interface point & Description \\
		\midrule
		Model specification & Construction and hyper-parameter specification \\
		Model estimation & Data ingestion and parameter estimation (e.g.\ \texttt{fit}) \\
		Application & Typically data transformation or generation of predictions (e.g.\ \texttt{predict}, \texttt{transform}) \\
		Updating & Online learning, e.g.\ \texttt{update} or \texttt{partial\_fit} in \software{scikit-learn} \\
		Inspection & While encapsulation will hide some information, key information such as hyper-parameters and fitted parameters should be accessible for users (e.g.\ \texttt{get\_params} and \texttt{get\_fitted\_params}) \\
		Model persistence & Saving fitted models for deployment (e.g.\ \texttt{save}) \\
		\bottomrule
	\end{tabularx}
\end{table}

In the following, we will discuss concrete examples, including supervised learners, transformers and composition algorithms such as pipelines and ensembles from scikit-learn, and forecasters and reduction algorithms from sktime.

\subsection{Supervised learners}

Following the ``consensus'' formulation of the supervised learning task introduced in section \ref{sec:cmodel.task}, we can define a scitype for a supervised learner as a combination of a class type specifying the interface and statistical properties. The class type can be define as follows:

\begin{table}[H]
	\begin{tabular}{l l c l}
		\multicolumn{4}{l}{\texttt{class type SupervisedLearner}}\\
		\quad \texttt{params} & \texttt{paramlist} & :& \texttt{paramobject}\\
		\quad \texttt{state} & \texttt{model}  & : & \texttt{mathobject}\\
		\quad \texttt{methods} & \texttt{fit} & : & $(\calX\times \calY)^N\rightarrow \texttt{model}$ \\
		\quad & \texttt{predict} & : &  $\calX \times \texttt{model} \rightarrow \calY$
	\end{tabular}
\end{table}

The cross-sectional nature of the input data, $\mathcal{X}$, implies two key statistical properties: (i) permutation invariability for the i.i.d.\ samples and (ii) feature permutation invariability. This means that, when represented in tabular form, the output of \texttt{fit} and \texttt{transform} should not change if the order of the rows (samples) and/or columns (features) is permuted.

Following the strategy/template pattern, the class type can serve as a blueprint for the abstract base class which other concrete implementations of supervised learners implement. Example \ref{code:sklearn-supervised} shows the implementation in \texttt{scikit-learn}.

\begin{minipage}{\textwidth}
\begin{lstlisting}[
caption=\software{scikit-learn}'s API for supervised learning (classification),
label=code:sklearn-supervised
]
classifier = RandomForestClassifier()
classifier.fit(y_train, X_train)
y_pred = classifier.predict(X_test)
\end{lstlisting}
\footnotesize{{\ttfamily X\_train} and {\ttfamily y\_train} denote the training data feature matrix and label vector, {\ttfamily X\_test} is the feature matrix of the test set, and {\ttfamily y\_pred} the predicted label.}
\end{minipage}\hfill

When comparing the \texttt{SupervisedLearner} scitype with scikit-learn's core API, it becomes clear that this scitype is central to \texttt{scikit-learn}'s interface. In particular, the universal base class of the \software{scikit-learn} toolbox is the \texttt{BaseEstimator} class. It implements some of our suggested universal interface points. For example, \texttt{get\_params} for parameter inspection, and generic functionality for construction and specification in the constructor (\texttt{\_\_init\_\_}).

We can, of course, further distinguish the \texttt{SupervisedLearner} class type into classifier and regressor types, according to the domain of the target variable, $\mathcal{Y}$. Classifiers and regressors are required to implement scitype-defining \texttt{fit} and \texttt{predict} methods with a uniform interface. In scikit-learn, there are also scitype specific base classes for regressors and classifiers, namely \texttt{RegressorMixin} and \texttt{ClassifierMixin}, but their remit is mainly accessory functionality (scoring) and scitype inspection (signposting by inheritance), rather than handling the \texttt{fit}/\texttt{predict} interface. As such, supervised classifiers and regressors in \texttt{scikit-learn} adhere to the template/strategy pattern. Note that, in scikit-learn, this pattern is enforced through conventions and unit testing, following ``design by contract'' principles \cite{Meyer1997}, rather than strict inheritance from abstract base classes.

Notably absent from \texttt{BaseEstimator} is a uniform interface for state variable inspections. Instead, inspection is specific to concrete classes and added on a descendant level. Instead, we argue that algorithms should also have a uniform interface for inspecting inference results (e.g.\ fitted coefficients) -- something which is currently missing from most major toolboxes. This would allow to develop composition classes which operate on fitted parameters of component algorithms.

\subsection{Forecasters}

Having illustrated how scitype-based patterns can motivate and explain core design choices in the existing toolboxes such as scikit-learn, we will now discuss how they can guide the design of new ones. We use the design of sktime's API for forecasting \cite{Loning2019, Loning2020} as an example.

We introduced the forecasting learning task in section \ref{sec:cmodel.task}. Following the same process as before, we can derive an algorithm scitype from the task specification. The class type can be specified as follows:

\begin{table}[H]
	\begin{tabular}{l l c l}
		\multicolumn{4}{l}{\texttt{class type Forecaster}}\\
		\quad \texttt{params} & \texttt{paramlist} & :& \texttt{paramobject}\\
		\quad \texttt{state} & \texttt{model}  & : & \texttt{mathobject}\\
		\quad \texttt{methods} & \texttt{fit} & : & $\mathcal{Z} \rightarrow$ \texttt{model} \\
		\quad & \texttt{predict} & : &  $\mathcal{T} \times \texttt{model} \rightarrow \mathcal{Z}$\\
	\end{tabular}
\end{table}
The time series nature of the input data implies two key statistical properties: (i) orderedness of time-series observations and (ii) shared domain of the observed time points in $\mathcal{Z}$ and forecasting horizon $\mathcal{T}$. This means that shuffling the observations may change the results of \texttt{fit}.

Example \ref{code:sktime-forecasting} shows the implementation in \software{sktime}.

\begin{minipage}{\textwidth}
	\small
\begin{lstlisting}[
caption=\software{sktime}'s API for (univariate) forecasting,
label=code:sktime-forecasting
]
forecaster = ExponentialSmoothing()
fh = [1, 2, 3]
forecaster.fit(y_train)
y_pred = forecaster.predict(fh)
\end{lstlisting}
	\footnotesize{{\ttfamily y\_train} denotes the training time series, {\ttfamily fh} the forecasting horizon for the three steps ahead relative to the end of the training series, and {\ttfamily y\_pred} the predicted values for the given forecasting horizon.}
\end{minipage}\hfill

Note that we pass the forecasting horizon, i.e.\ task information, explicitly to the algorithm. In the previous example from scikit-learn, we provide necessary task information, such as the target variable, by through separation of input arguments, i.e.\ we pass two arguments, one for the target variable and one for the feature data to the \texttt{fit} method, with \texttt{y\_train} being the designated target variable and \texttt{X\_train} the feature data. By contrast, in this example, we define a separate \texttt{fh} object to pass that information to the forecasting algorithm following the co-strategy pattern in section \ref{sec:patterns.ML}. Both designs adhere to the design principle of keeping the specification of task information -- i.e.\ the target name or forecasting horizon -- separate from specification of the algorithm as described in section \ref{sec:principles.layers}.

Also note that in contrast to the tabular (or cross-sectional) setting, there is no established standard interface design for the forecasting setting, or other related time series learning problems. Instead, the ecosystem is fragmented with various specialized toolkits for specific algorithm families \cite{Loning2019}. We hope that our design principles and patterns presented here can guide future toolbox development and enable us to provide advanced time series toolbox capabilities.

\subsection{Composition}

Until now, we have discussed primitive (or atomic) learning algorithms. In this section, we turn to composite algorithms. Composite algorithms, sometimes also called meta-algorithms, consist of and operate on one or more component algorithms. Common examples include tuning, pipelining or ensembling.

\subsubsection{Pipelining, ensembling and tuning}
In section \ref{sec:patterns.ML}, we outline the key composition patterns for ML toolboxes. Scitypes extend directly to higher-order scitypes to describe composite algorithms. For example, one of most common composite algorithms is a pipeline which, in its simplest form, chains multiple prior data transformations with a final learning algorithm. As a higher-order scitype, this can be expressed as:

\texttt{Pipeline}: $(\texttt{Transformer})^n \times \texttt{SupervisedLearner} \rightarrow \texttt{SupervisedLearner}$,

where a pipeline is defined as taking $n$ transformers and a final supervised learner and composing them into a single, composite supervised learner. Similarly, we can define an ensemble of supervised learners:

\texttt{Ensemble}: $(\texttt{SupervisedLearner})^n \rightarrow \texttt{SupervisedLearner}$.

The composite algorithm adhere to the scitype composition pattern as described in section \ref{sec:patterns.ML}. As before, we can follow the scitype template/strategy pattern and encapsulate each higher-order scitype as an abstract base class, using the \texttt{Pipeline} scitype together with the \texttt{SupervisedLearner} as a blueprint for an abstract base class.

The key advantage of the pattern is that composite and individual algorithms can be treated uniformly. Primitive and composite object become interchangeable at run time. For example, a pipeline has the same scitype as its final learning algorithm. Likewise, the ensemble is of the same scitype as the component algorithms. As a consequence, users (and other client algorithms) can ignore the difference between compositions of algorithms and individual algorithms. For example, in \texttt{scikit-learn}, tuning procedure \texttt{GridSearchCV} and pipelining operation \texttt{Pipeline} adhere to the composite pattern.

We make a few additional remarks:
\begin{itemize}
	\item We encapsulate composition algorithms in separate classes instead of assigning additional methods to the base class mainly for two reasons: First, composition approaches are not part of the simplest possible representation of scitypes following the principle laid out in section \ref{sec:principles.scitype}. Second, there are often different concrete variants of composition algorithms. For example there are different algorithms for tuning (e.g.\ full grid search, randomized grid search, successive halving). Encapsulating composition algorithm in separate classes makes it easy to add new meta algorithms without changing any other class. In addition, many composition learning algorithm are often general and not closely tied to specific learning algorithms. They are often applicable to any algorithm of the same scitype, however, there are some specialized approaches closely connected to concrete algorithm implementations which are implemented in separate classes.
	\item Composite algorithms require consistent interfaces as imposed by the strategy/template pattern. This is because the composite algorithm usually forwards a method call to its component algorithms, and may  performs additional operations before and/or after forwarding. For example, when ensembling algorithms, the composite algorithm will forward the fit call to all individual ensemble members, when calling predict, it will forward the predict call to all ensemble members and then return an aggregated prediction. The forwarding only works if the composition class can expect a consistent interface.
	\item In order for ML composition to work well, composition must allow nested hyper-parameter and parameter access. Hyper-parameters of component algorithms must be accessible and settable from the interface of the composite algorithm. 
	\item Compositions may also make use of the decorator pattern from \citet{Gamma2002}. In this context, this pattern allows to assign responsibilities to an object dynamically at run time, depending on its component models. For example, a composite algorithm may expose some methods or parameters from its component algorithms, but the exact methods and fields it exposes depend on the type of component algorithm specified at run time. For example, this is implemented in \software{scikit-learn} in the \texttt{GridSearchCV} class.
\end{itemize}

\subsubsection{Transformers}
Data transformations are commonly encapsulated as transformers. The main reason for encapsulating transformations in transformer classes is that we do not keep track of data changes in the data object itself, which is usually conceived of as an immutable value object (see section \ref{sec:cmodel}). Instead, we treat data transformations as part of model specification. Note that unlike the learning algorithms discussed so far, transformers by themselves do not correspond to any learning task. We can, however, still define scitypes for transformations, based on their key operations and statistical properties in the context of learning tasks that they are applied to. In terms of scitypes, we can distinguish ``primary scitypes'' that are directly addressing a formal task and ``secondary scitypes'' which arise as operations related to primary scitypes (e.g.\ transformers). For example, viewed in the context of supervised learning pipelines, transformers need to be compatible with the established scitype for supervised learners, i.e.\ separate methods for fitting, application of the fitted model (\texttt{predict} and \texttt{transform} respectively), as well as the common hyper-parameter interface. In this sense, transformers can be conceived of as modifications to supervised learners together with the pipelines that they are part of.

We can see how the transformer API in \software{scikit-learn} implements the scitype.

\begin{table}[H]
	\begin{tabular}{l l c l}
	\multicolumn{4}{l}{\texttt{class type Transformer}}\\
	\quad \texttt{params} & \texttt{paramlist} & :& \texttt{paramobject}\\
	\quad \texttt{state} & \texttt{model}  & : & \texttt{mathobject}\\
	\quad \texttt{methods} & \texttt{fit} & : & $(\calX\times \calY)^N\rightarrow \texttt{model}$ \\
	\quad & \texttt{predict} & : &  $\calX \times \texttt{model} \rightarrow \mathcal{U}$
\end{tabular}
\end{table}

Note that this is a cross-sectional transformer scitype, given the cross-sectional domain of the input data $\mathcal{X}$ and transformed output data $\mathcal{U}$. As a consequence, the same statistical properties hold as for the \texttt{SupervisedLearner} scitype. In principle, transformer scitypes for other data formats are possible, for example for time series data (e.g.\ Box-Cox transformations). Example \ref{code:sklearn-transformers} shows the implementation in \software{scikit-learn}.

\begin{minipage}{\textwidth}
\begin{lstlisting}[
	caption=\software{scikit-learns}'s API for transformers,
	label=code:sklearn-transformers
	]
	transformer = PCA()
	transformer.fit(X, y)
	Xt = transformer.transform(X)
\end{lstlisting}
\footnotesize{{\ttfamily X} and {\ttfamily y} denote the some cross-sectional data and target vector, and {\ttfamily Xt} the transformed data.}
\end{minipage}\hfill

\subsubsection{Reduction}
The final example we discuss is that of reduction. Note that reduction sometimes also refers to dimensionality reduction. We here instead refer to reduction as an approach to convert one learning task into another one, allowing us to apply an algorithm for one task to solve another task \cite{Beygelzimer2015, Beygelzimer2005a}. A classical example in supervised learning is one-vs-all classification, reducing k-way multi-category classification to k binary classification tasks.

Another common example is reducing forecasting to supervised regression \citet{Bontempi2012}. This typically involves the transformation of the time series data that one wants to forecast into the required tabular input format, so that one can then fit and apply any regression algorithm. The higher-order scitype for this particular reduction approach can be expressed as follows:

\texttt{ForecastingToRegressionReduction}: $\texttt{SupervisedLearner} \rightarrow \texttt{Forecaster}$

where the composition algorithm takes in any algorithm of the scitype \texttt{SupervisedLearner} and adapts it to the interface of the \texttt{Forecaster} scitype.

While reductions are not new, we are the first to propose encapsulating them as meta-estimators in a separate class following the composition pattern in section \ref{sec:patterns.ML}. Reductions have several key properties that make them well suited to be expressed as meta-estimators:
\begin{itemize}
	\litem{Modularity.} Reductions convert any algorithm for a particular task into an algorithm for a new task. Applying some reduction approach to $n$ base algorithms gives $n$ new algorithms for the new task. Any progress on the base algorithm immediately transfers to the new task, saving both research and software development effort \cite{Beygelzimer2005, Beygelzimer2008}.
	\litem{Tunability.} Most reductions require modeling choices that we may want to optimize. For example, we may want to tune the window length or select among different strategies for generating forecasts (e.g.\ ``direct'' or ``recursive'') \cite{Taieb2014, Bontempi2012}. By expressing reductions as meta-estimators, we expose these choices via the common interface as tunable hyper-parameters.
	\litem{Composability.} Reductions are composable. They can be composed to solve more complicated problems \cite{Beygelzimer2005, Beygelzimer2008}. For example, we can first reduce forecasting to time series regression which in turn can be reduced to tabular regression via feature extraction.
	\litem{Adaptability.} Reductions are adapters. They adapt the interface of the base algorithm to the interface required for solving the new task, following the adapter pattern in \citet{Gamma2002} and the scitype reduction pattern in section \ref{sec:patterns.ML}. This lets users apply algorithms to learning tasks that could not otherwise not be applied to those tasks because of incompatible interfaces and to reuse the common tuning and model evaluation tools appropriate for the new task. 
\end{itemize}

\section{Conclusion}
\label{sec:conclusion}

ML toolboxes have become the workhorses of modern data scientific practice. However, despite their universal success, the key principles in their design have never been fully analyzed.

In this paper, we presented a first attempt at analyzing ML toolbox design. Following a domain-driven design methodology, we first developed a conceptual model for common objects in the ML/AI domain. At its core, we proposed a new type system, called scientific typing, which captures the data scientific purpose and meaning of based on the set of operations that we usually perform with them (i.e.\ their interface) and their statistical properties. The proposed conceptual model is both well-grounded in the underlying mathematical and statistical formalism and easily translatable into software.

In our analysis, we combined ideas from classical software engineering, type theory and formal mathematical statistics, as well as insights from established toolboxes in the ML/AI domain. From our analysis, we derived a set of key design principles and reusable patterns using object-oriented programming. We illustrated that these principles and patterns can not only explain the design of existing toolboxes such as scikit-learn, but also guide the development of new ones.

We hope that our contribution can serve as a state-of-the-art reference for ML practitioners and developers, and, potentially, as a first step towards a higher-level declarative language for constructing AI.

\section*{Acknowledgments}
We would like to thank Viktor Kazakov for his feedback and discussions on a first draft of the paper. We would also like to thank the contributor communities of the \software{distr6}, \software{mlr3proba}, \software{MLJ} and \software{sktime} projects for their indirect contributions and feedback.

Markus Löning's contribution was supported by the UK Economic and Social Research Council (ESRC grant: ES-P000592-1), the Consumer Data Research Centre (CDRC) (ESRC grant: ES-L011840-1), and The Alan Turing Institute (EPSRC grant: EP-N510129-1).

\section*{Authors' contributions}
FK initiated the project and provided key ideas. All authors contributed to the development of ideas, primarily through collaboration in open source projects: \software{distr6}, \software{mlr3}, \software{mlaut}, \software{MLJ}, \software{sktime}, and AG's MSc thesis (\software{pysf}). AB introduced the term ``scitype'' and the formal idea in the context of data containers and the \software{MLJ} package. FK conceived the concept of abstract scitypes of formal objects beyond data containers. FK and ML led and coordinated the writing effort, with all authors contributing at least in an editor role. All authors are jointly responsible for content and exposition.

\bibliographystyle{plainnat}
\bibliography{references}

\begin{thebibliography}{45}
\providecommand{\natexlab}[1]{#1}
\providecommand{\url}[1]{\texttt{#1}}
\expandafter\ifx\csname urlstyle\endcsname\relax
  \providecommand{\doi}[1]{doi: #1}\else
  \providecommand{\doi}{doi: \begingroup \urlstyle{rm}\Url}\fi

\bibitem[Abadi et~al.(2016)Abadi, Barham, Chen, Chen, Davis, Dean, Devin,
  Ghemawat, Irving, Isard, and Others]{Abadi2016}
Martin Abadi, Paul Barham, Jianmin Chen, Zhifeng Chen, Andy Davis, Jeffrey
  Dean, Matthieu Devin, Sanjay Ghemawat, Geoffrey Irving, Michael Isard, and
  Others.
\newblock {TensorFlow: A system for large-scale machine learning}.
\newblock In \emph{12th USENIX Symposium on Operating Systems Design and
  Implementation (OSDI 16)}, pages 265--283, 2016.

\bibitem[Bass et~al.(2003)Bass, Clements, and Kazman]{Bass2003}
Len Bass, Paul Clements, and Rick Kazman.
\newblock \emph{{Software architecture in practice}}.
\newblock Addison-Wesley Professional, 2003.

\bibitem[Beygelzimer et~al.(2005{\natexlab{a}})Beygelzimer, Dani, Hayes,
  Langford, and Zadrozny]{Beygelzimer2005a}
Alina Beygelzimer, Varsha Dani, Tom Hayes, John Langford, and Bianca Zadrozny.
\newblock {Error limiting reductions between classification tasks}.
\newblock In \emph{Proceedings of the 22nd international conference on Machine
  learning}, pages 49--56. ACM, 2005{\natexlab{a}}.

\bibitem[Beygelzimer et~al.(2005{\natexlab{b}})Beygelzimer, Langford, and
  Zadrozny]{Beygelzimer2005}
Alina Beygelzimer, John Langford, and Bianca Zadrozny.
\newblock {Weighted one-against-all}.
\newblock In \emph{American Association for Artificial Intelligence (AAAI)},
  pages 720--725, 2005{\natexlab{b}}.

\bibitem[Beygelzimer et~al.(2008)Beygelzimer, Langford, and
  Zadrozny]{Beygelzimer2008}
Alina Beygelzimer, John Langford, and Bianca Zadrozny.
\newblock {Machine learning techniques—reductions between prediction quality
  metrics}.
\newblock In \emph{Performance Modeling and Engineering}, pages 3--28.
  Springer, 2008.

\bibitem[Beygelzimer et~al.(2015)Beygelzimer, Daum{\'{e}}, Langford, and
  Mineiro]{Beygelzimer2015}
Alina Beygelzimer, Hal Daum{\'{e}}, John Langford, and Paul Mineiro.
\newblock {Learning reductions that really work}.
\newblock \emph{Proceedings of the IEEE}, 104\penalty0 (1):\penalty0 136--147,
  2015.

\bibitem[Bischl et~al.(2016)Bischl, Lang, Kotthoff, Schiffner, Richter,
  Studerus, Casalicchio, and Jones]{Bischl2016}
Bernd Bischl, Michel Lang, Lars Kotthoff, Julia Schiffner, Jakob Richter, Erich
  Studerus, Giuseppe Casalicchio, and Zachary~M. Jones.
\newblock {mlr: Machine Learning in R}.
\newblock \emph{Journal of Machine Learning Research}, 17\penalty0
  (170):\penalty0 1--5, 2016.
\newblock URL \url{http://jmlr.org/papers/v17/15-066.html}.

\bibitem[Blaom et~al.(2020{\natexlab{a}})Blaom, Lienart, Simillides, Arenas,
  Vollmersj, Giordano, Samuel, Shridhar, Shridhar, Ed, Swenkel, Samaroo,
  Evalparse, Hoffimann, Sjvollmer, Borregaard, Squire, Pshashk, Lhnguyen-vn,
  Azev77, Agrawal, Venkateshprasad, H{\"{o}}nig, Nils, Kryohi, TagBot,
  Gabasova, Aluthge, and St-Jean]{Blaom2020}
Anthony Blaom, Thibaut Lienart, Yiannis Simillides, Diego Arenas, Vollmersj,
  Mos{\`{e}} Giordano, Okon Samuel, Ayush Shridhar, Ayush Shridhar, Ed,
  Swenkel, Julian Samaroo, Evalparse, J{\'{u}}lio Hoffimann, Sjvollmer,
  Michael~Krabbe Borregaard, Kevin Squire, Pshashk, Lhnguyen-vn, Azev77, Ashrya
  Agrawal, Venkateshprasad, Robert H{\"{o}}nig, Nils, Kryohi, Julia TagBot,
  Evelina Gabasova, Dilum Aluthge, and C{\'{e}}dric St-Jean.
\newblock {MLJ: A Machine Learning Framework for Julia}, apr
  2020{\natexlab{a}}.
\newblock URL \url{https://zenodo.org/record/3765808}.

\bibitem[Blaom et~al.(2020{\natexlab{b}})Blaom, Kiraly, Lienart, Simillides,
  Arenas, and Vollmer]{Blaom2020a}
Anthony~D Blaom, Franz Kiraly, Thibaut Lienart, Yiannis Simillides, Diego
  Arenas, and Sebastian~J Vollmer.
\newblock {MLJ: A Julia package for composable machine learning}.
\newblock \emph{Journal of Open Source Software}, 5\penalty0 (55):\penalty0
  2704, 2020{\natexlab{b}}.
\newblock \doi{10.21105/joss.02704}.
\newblock URL \url{https://doi.org/10.21105/joss.02704}.

\bibitem[Bontempi et~al.(2013)Bontempi, Taieb, and {Le Borgne}]{Bontempi2012}
Gianluca Bontempi, Souhaib~Ben Taieb, and Yann-A{\"{e}}l {Le Borgne}.
\newblock {Machine Learning Strategies for Time Series Forecasting}.
\newblock In \emph{Business Intelligence}, pages 62--77. Springer, Berlin,
  Heidelberg, 2013.

\bibitem[Box et~al.(2013)Box, Jenkins, and Reinsel]{Box2013}
George E.~P. Box, Gwilym~M. Jenkins, and Gregory~C. Reinsel.
\newblock \emph{{Time series analysis: Forecasting and control: Fourth
  edition}}.
\newblock 2013.
\newblock ISBN 9781118619193.
\newblock \doi{10.1002/9781118619193}.

\bibitem[Brown et~al.(1998)Brown, Malveau, Mowbray, and Wiley]{brown1998anti}
William~J Brown, Raphael~C Malveau, Thomas~J Mowbray, and John Wiley.
\newblock \emph{{AntiPatterns: Refactoring Software , Architectures, and
  Projects in Crisis}}.
\newblock Wiley, 1998.
\newblock ISBN 0849329949.

\bibitem[Buitinck et~al.(2013)Buitinck, Louppe, Blondel, Pedregosa,
  M{\"{u}}ller, Grisel, Niculae, Prettenhofer, Gramfort, Grobler, Layton,
  Vanderplas, Joly, Holt, and Varoquaux]{Buitinck2013}
Lars Buitinck, Gilles Louppe, Mathieu Blondel, Fabian Pedregosa, Andreas~C
  M{\"{u}}ller, Olivier Grisel, Vlad Niculae, Peter Prettenhofer, Alexandre
  Gramfort, Jaques Grobler, Robert Layton, Jake Vanderplas, Arnaud Joly, Brian
  Holt, and Ga{\"{e}}l Varoquaux.
\newblock {API design for machine learning software: experiences from the
  scikit-learn project}.
\newblock In \emph{ECML PKDD 2013 Workshop on Languages for Data Mining and
  Machine Learning}, 2013.
\newblock URL \url{https://github.com/scikit-learn}.

\bibitem[Buschmann et~al.(2007)Buschmann, Henney, and Schimdt]{Buschmann2007}
Frank Buschmann, Kelvin Henney, and Douglas Schimdt.
\newblock \emph{{Pattern-Oriented Software Architecture: On Patterns And
  Pattern Language, Volume 5}}, volume~5.
\newblock John wiley {\&} sons, 2007.

\bibitem[Chambers and Others(2014)]{Chambers2014}
John~M Chambers and Others.
\newblock {Object-oriented programming, functional programming and R}.
\newblock \emph{Statistical Science}, 29\penalty0 (2):\penalty0 167--180, 2014.

\bibitem[Evans(2004)]{Evans2004}
Eric Evans.
\newblock \emph{{Domain-driven design: tackling complexity in the heart of
  software}}.
\newblock Addison-Wesley Professional, 2004.

\bibitem[Gamma et~al.(1997)Gamma, Helm, Johnson, and Vlissides]{Gamma2002}
Erich Gamma, Richard Helm, Ralph Johnson, and John Vlissides.
\newblock \emph{{Design Patterns – Elements of Reusable Object-Oriented
  Software}}.
\newblock Addison Wesley Longman, Inc., 1997.
\newblock ISBN 9780201715941.
\newblock \doi{10.1093/carcin/bgs084}.

\bibitem[Gressmann and Kir{\'{a}}ly(2018)]{Gressmann2018a}
Frithjof Gressmann and Franz~J. Kir{\'{a}}ly.
\newblock {skpro: A domain-agnostic modelling framework for probabilistic
  supervised learning}.
\newblock \emph{openreview.net}, 2018.
\newblock URL \url{https://openreview.net/forum?id=B1gIZHtAFQ}.

\bibitem[Guecioueur(2018)]{Guecioueur2018}
Ahmed Guecioueur.
\newblock {pysf: Supervised forecasting of sequential data in Python}, 2018.
\newblock URL \url{https://pypi.org/project/pysf/}.

\bibitem[Hall et~al.(2009)Hall, Frank, Holmes, Pfahringer, Reutemann, and
  Witten]{Hall2009}
Mark Hall, Eibe Frank, Geoffrey Holmes, Bernhard Pfahringer, Peter Reutemann,
  and Ian~H. Witten.
\newblock {The WEKA data mining software}.
\newblock \emph{ACM SIGKDD Explorations Newsletter}, 11\penalty0 (1):\penalty0
  10, nov 2009.
\newblock ISSN 19310145.
\newblock \doi{10.1145/1656274.1656278}.
\newblock URL \url{http://portal.acm.org/citation.cfm?doid=1656274.1656278}.

\bibitem[Hastie et~al.(2009)Hastie, Tibshirani, and Friedman]{Hastie2009}
Trevor Hastie, Robert~T. Tibshirani, and Jerome Friedman.
\newblock \emph{{The Elements of Statistical Learning}}, volume~1.
\newblock Springer, 2 edition, 2009.
\newblock ISBN 978-0-387-84857-0.
\newblock \doi{10.1007/b94608}.
\newblock URL \url{http://www.springerlink.com/index/10.1007/b94608}.

\bibitem[Kazakov and Kir{\'{a}}ly(2019)]{Kazakov2019}
Viktor Kazakov and Franz~J Kir{\'{a}}ly.
\newblock {Machine Learning Automation Toolbox (MLaut)}.
\newblock \emph{arXiv preprint arXiv:1901.03678}, 2019.

\bibitem[Kuhn(2008)]{Kuhn2008}
Max Kuhn.
\newblock {Building Predictive Models in R: Using the caret Package}.
\newblock \emph{Journal of Statistical Software}, 28\penalty0 (5), 2008.
\newblock \doi{10.18637/jss.v028.i05}.
\newblock URL \url{http://www.jstatsoft.org/v28/i05/}.

\bibitem[Lang et~al.(2019)Lang, Binder, Richter, Schratz, Pfisterer, Coors, Au,
  Casalicchio, Kotthoff, and Bischl]{lang2019mlr3}
Michel Lang, Martin Binder, Jakob Richter, Patrick Schratz, Florian Pfisterer,
  Stefan Coors, Quay Au, Giuseppe Casalicchio, Lars Kotthoff, and Bernd Bischl.
\newblock {mlr3: A modern object-oriented machine learning framework in R}.
\newblock \emph{Journal of Open Source Software}, 4\penalty0 (44):\penalty0
  1903, 2019.

\bibitem[Larman(2012)]{Larman2012}
Craig Larman.
\newblock \emph{{Applying UML and patterns: an introduction to object oriented
  analysis and design and interative development}}.
\newblock Pearson Education India, 2012.

\bibitem[L{\"{o}}ning and Kir{\'{a}}ly(2020)]{Loning2020}
Markus L{\"{o}}ning and Franz~J. Kir{\'{a}}ly.
\newblock {Forecasting with sktime: Designing sktime's New Forecasting API and
  Applying It to Replicate and Extend the M4 Study}.
\newblock \emph{ArXiv e-prints}, 2020.

\bibitem[L{\"{o}}ning et~al.(2019)L{\"{o}}ning, Bagnall, Ganesh, Kazakov,
  Lines, and Kir{\'{a}}ly]{Loning2019}
Markus L{\"{o}}ning, Anthony Bagnall, Sajaysurya Ganesh, Viktor Kazakov, Jason
  Lines, and Franz~J Kir{\'{a}}ly.
\newblock {sktime: A Unified Interface for Machine Learning with Time Series}.
\newblock \emph{Workshop on Systems for ML at NeurIPS 2019}, 2019.

\bibitem[Meyer(1997)]{Meyer1997}
Bertrand Meyer.
\newblock \emph{{Object-oriented software construction}}, volume~2.
\newblock Prentice hall Englewood Cliffs, 1997.

\bibitem[Nalchigar et~al.(2019)Nalchigar, Yu, Obeidi, Carbajales, Green, and
  Chan]{nalchigar2019solution}
Soroosh Nalchigar, Eric Yu, Yazan Obeidi, Sebastian Carbajales, John Green, and
  Allen Chan.
\newblock {Solution patterns for machine learning}.
\newblock In \emph{International Conference on Advanced Information Systems
  Engineering}, pages 627--642. Springer, 2019.

\bibitem[Nascimento et~al.(2020)Nascimento, Nguyen-Duc, Sundb{\o}, and
  Conte]{nascimento2020software}
Elizamary Nascimento, Anh Nguyen-Duc, Ingrid Sundb{\o}, and Tayana Conte.
\newblock {Software engineering for artificial intelligence and machine
  learning software: A systematic literature review}.
\newblock \emph{arXiv preprint arXiv:2011.03751}, 2020.

\bibitem[Paszke et~al.(2019)Paszke, Gross, Massa, Lerer, Bradbury, Chanan,
  Killeen, Lin, Gimelshein, Antiga, and Others]{Paszke2019}
Adam Paszke, Sam Gross, Francisco Massa, Adam Lerer, James Bradbury, Gregory
  Chanan, Trevor Killeen, Zeming Lin, Natalia Gimelshein, Luca Antiga, and
  Others.
\newblock {Pytorch: An imperative style, high-performance deep learning
  library}.
\newblock In \emph{Advances in neural information processing systems}, pages
  8026--8037, 2019.

\bibitem[Pedregosa et~al.(2011)Pedregosa, Varoquaux, Gramfort, Michel, Thirion,
  Grisel, Blondel, Prettenhofer, Weiss, Dubourg, Vanderplas, Passos,
  Cournapeau, Brucher, Perrot, and Duchesnay]{Pedregosa2001}
Fabian Pedregosa, Ga{\"{e}}l Varoquaux, Alexandre Gramfort, Vincent Michel,
  Bertrand Thirion, Olivier Grisel, Mathieu Blondel, Peter Prettenhofer, Ron
  Weiss, Vincent Dubourg, Jake Vanderplas, Alexandre Passos, David Cournapeau,
  Matthieu Brucher, Matthieu Perrot, and Edouard Duchesnay.
\newblock {Scikit-learn: Machine Learning in Python}.
\newblock \emph{The Journal of Machine Learning Research}, 12:\penalty0
  2825--2830, 2011.
\newblock URL \url{https://dl.acm.org/citation.cfm?id=2078195}.

\bibitem[Pierce and Benjamin(2002)]{Pierce2002}
Benjamin~C Pierce and C~Benjamin.
\newblock \emph{{Types and programming languages}}.
\newblock 2002.

\bibitem[Pressman(2005)]{Pressman2005}
Roger~S Pressman.
\newblock \emph{{Software engineering: a practitioner's approach}}.
\newblock Palgrave macmillan, 2005.

\bibitem[{R Core Team}(2014)]{RCoreTeam2014}
{R Core Team}.
\newblock {R: A Language and Environment for Statistical Computing}.
\newblock Technical report, R Foundation for Statistical Computing, Vienna,
  Austria, 2014.

\bibitem[Sculley et~al.(2015)Sculley, Holt, Golovin, Davydov, Phillips, Ebner,
  Chaudhary, Young, Crespo, and Dennison]{sculley2015hidden}
David Sculley, Gary Holt, Daniel Golovin, Eugene Davydov, Todd Phillips,
  Dietmar Ebner, Vinay Chaudhary, Michael Young, Jean-Francois Crespo, and Dan
  Dennison.
\newblock {Hidden technical debt in machine learning systems}.
\newblock In \emph{Advances in neural information processing systems}, pages
  2503--2511, 2015.

\bibitem[Sonabend et~al.(2020)Sonabend, Kir{\'{a}}ly, Bender, Bischl, and
  Lang]{Sonabend2020}
Raphael Sonabend, Franz~J Kir{\'{a}}ly, Andreas Bender, Bernd Bischl, and
  Michel Lang.
\newblock {mlr3proba: Machine Learning Survival Analysis in R}.
\newblock \emph{arXiv preprint arXiv:2008.08080}, 2020.

\bibitem[Sonnenburg et~al.(2007)Sonnenburg, Braun, Ong, Bengio, Bottou, Holmes,
  LeCun, M{\"{u}}ller, Pereira, Rasmussen, and Others]{Sonnenburg2007}
S{\"{o}}ren Sonnenburg, Mikio~L Braun, Cheng~Soon Ong, Samy Bengio, Leon
  Bottou, Geoffrey Holmes, Yann LeCun, Klaus-Robert M{\"{u}}ller, Fernando
  Pereira, Carl~Edward Rasmussen, and Others.
\newblock {The need for open source software in machine learning}.
\newblock \emph{Journal of Machine Learning Research}, 8\penalty0
  (Oct):\penalty0 2443--2466, 2007.

\bibitem[Taieb(2014)]{Taieb2014}
Souhaib~Ben Taieb.
\newblock \emph{{Machine learning strategies for multi-step-ahead time series
  forecasting}}.
\newblock PhD thesis, Universit Libre de Bruxelles, Belgium, 2014.

\bibitem[Vanschoren et~al.(2014)Vanschoren, {Van Rijn}, Bischl, and
  Torgo]{Vanschoren2014}
Joaquin Vanschoren, Jan~N {Van Rijn}, Bernd Bischl, and Luis Torgo.
\newblock {OpenML: networked science in machine learning}.
\newblock \emph{ACM SIGKDD Explorations Newsletter}, 15\penalty0 (2):\penalty0
  49--60, 2014.

\bibitem[Varoquaux et~al.(2015)Varoquaux, Buitinck, Louppe, Grisel, Pedregosa,
  and Mueller]{Varoquaux2015}
Ga{\"{e}}l Varoquaux, Lars Buitinck, Gilles Louppe, Olivier Grisel,
  F.~Pedregosa, and A.~Mueller.
\newblock {Scikit-learn: Machine Learning Without Learning the Machinery}.
\newblock \emph{GetMobile: Mobile Computing and Communications}, 19\penalty0
  (1):\penalty0 29--33, jun 2015.
\newblock \doi{10.1145/2786984.2786995}.
\newblock URL \url{http://dl.acm.org/citation.cfm?doid=2786984.2786995}.

\bibitem[Wirfs-Brock and McKean(2003)]{Wirfs-Brock2003}
Rebecca Wirfs-Brock and Alan McKean.
\newblock \emph{{Object design: roles, responsibilities, and collaborations}}.
\newblock Addison-Wesley Professional, 2003.

\bibitem[Wirfs-Brock and Wilkerson(1989)]{Wirfs-Brock1989}
Rebecca Wirfs-Brock and Brian Wilkerson.
\newblock {Object-oriented design: a responsibility-driven approach}.
\newblock \emph{ACM sigplan notices}, 24\penalty0 (10):\penalty0 71--75, 1989.

\bibitem[Wirfs-Brock et~al.(1990)Wirfs-Brock, Wilkerson, and
  Wiener]{Wirfs-Brock1990}
Rebecca Wirfs-Brock, Brian Wilkerson, and Lauren Wiener.
\newblock {Designing object-oriented software}.
\newblock 1990.

\bibitem[Zheng(2014)]{Zheng2014}
A~Zheng.
\newblock {The challenges of building machine learning tools for the masses}.
\newblock In \emph{SE4ML:SoftwareEngineer- ing for Machine Learning (NeurIPS
  2014 Workshop)}, 2014.

\end{thebibliography}

\end{document}